\documentclass[review]{elsarticle}

\usepackage{graphicx}
\usepackage{caption}
\usepackage{subcaption}

\usepackage{lineno,hyperref,listings,xspace,xcolor}
\usepackage{xcolor}
\definecolor{codegreen}{rgb}{0,0.6,0}
\definecolor{codegray}{rgb}{0.5,0.5,0.5}
\definecolor{codepurple}{rgb}{0.58,0,0.82}
\definecolor{backcolour}{rgb}{0.95,0.95,0.92}

\lstdefinestyle{mystyle}{
    backgroundcolor=\color{backcolour},   
    commentstyle=\color{codegreen},
    keywordstyle=\color{magenta},
    numberstyle=\tiny\color{codegray},
    stringstyle=\color{codepurple},
    basicstyle=\ttfamily\footnotesize,
    breakatwhitespace=false,         
    breaklines=true,                 
    captionpos=b,                    
    keepspaces=true,                 
    numbers=left,                    
    numbersep=5pt,                  
    showspaces=false,                
    showstringspaces=false,
    showtabs=false,                  
    tabsize=2
}
\journal{Astronomy \& Computing}

\biboptions{authoryear}

\newcommand*{\astorb}{\texttt{astorb}\xspace}

\begin{document}

\begin{frontmatter}

\title{The \astorb database at Lowell Observatory}

\author[lowell]{Nicholas A. Moskovitz}
\cortext[mycorrespondingauthor]{Corresponding author}
\ead{nmosko@lowell.edu}

\author[lowell]{Lawrence Wasserman}
\author[lowell]{Brian Burt}
\author[lowell]{Robert Schottland}
\author[lowell]{Edward Bowell}
\author[usgs]{Mark Bailen}
\author[helsinki,lulea]{Mikael Granvik}

\address[lowell]{Lowell Observatory, 1400 W Mars Hill Road, Flagstaff, AZ 86001, USA}
\address[usgs]{USGS Astrogeology, Flagstaff, AZ}
\address[helsinki]{Department of Physics, P.O. Box 64, 00014 University of Helsinki, Finland}
\address[lulea]{Asteroid Engineering Lab, Lule\r{a}  University of Technology, Box 848, SE-981 28 Kiruna, Sweden}

\begin{abstract}

The \astorb database at Lowell Observatory is an actively curated catalog of all known asteroids in the Solar System. \astorb has heritage dating back to the 1970's and has been publicly accessible since the 1990's. Beginning in 2015 work began to modernize the underlying database infrastructure, operational software, and associated web applications. That effort has involved the expansion of \astorb to incorporate new data such as physical properties (e.g. albedo, colors, spectral types) from a variety of sources. The data in \astorb are used to support a number of research tools hosted at \url{https://asteroid.lowell.edu}. Here we present a full description of the software tools, computational foundation, and data products upon which the \astorb ecosystem has been built.

\end{abstract}


\end{frontmatter}

\section{The history of \astorb} \label{sec:history}

\begin{figure}[h]
    \includegraphics[width=\textwidth]{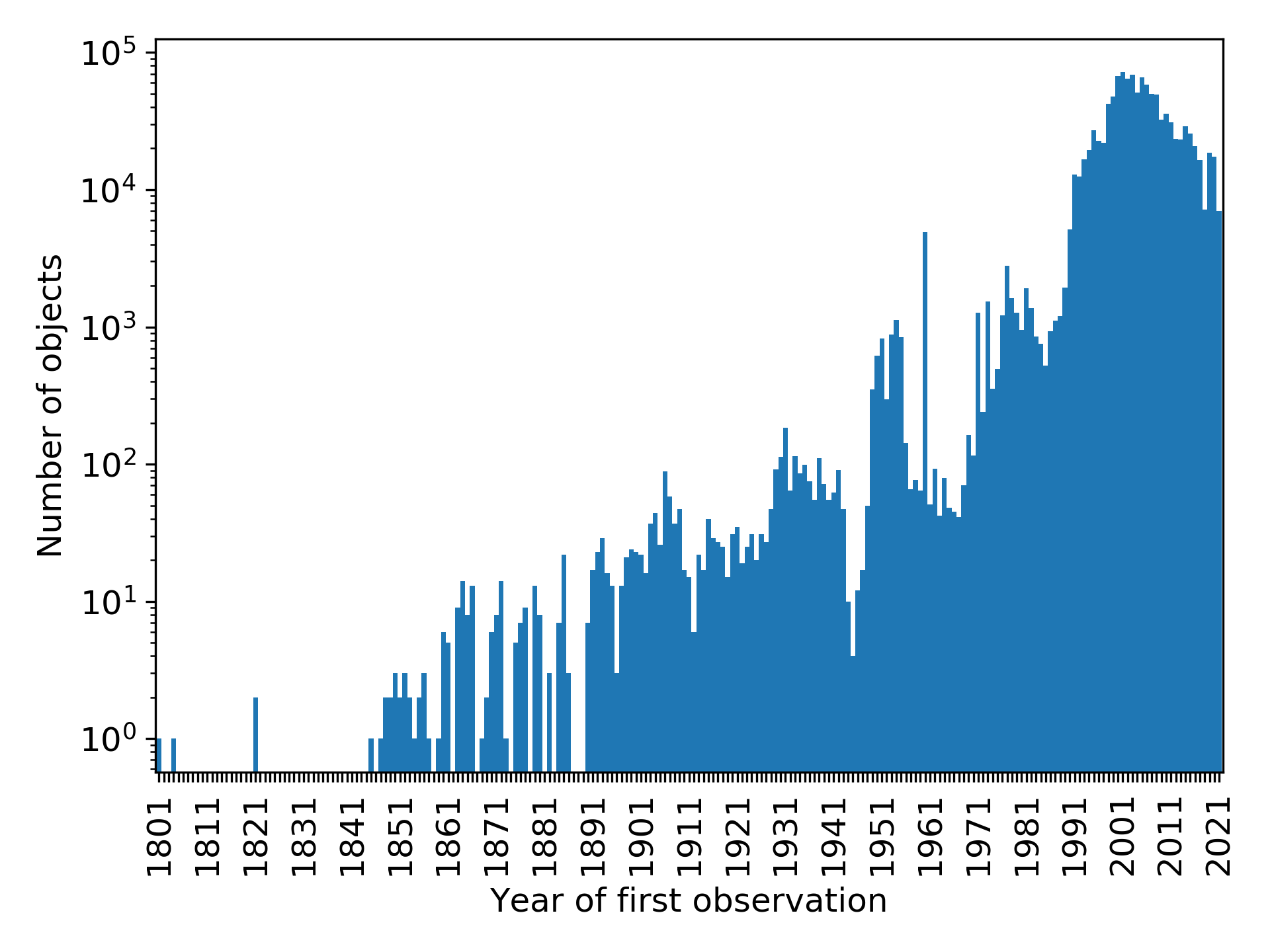}
    \caption{Histogram of first observation dates for all asteroids in \astorb as of 25 July 2021. The dramatic growth in the number of known minor planets will continue as next generation surveys like the Vera Rubin Observatory's Legacy Survey of Space and Time (LSST) begin in the mid 2020's \citep{Jones09}.}
    \label{fig:firstobs}
\end{figure}

Over the past 50 years the number of known minor planets in the Solar System has increased by 2.5 orders of magnitude from just under 4000 objects in 1970 to more than 1.2 million in 2022 (Figure \ref{fig:firstobs}). Maintaining catalogs of minor planets has required increasing effort from the few organizations around the world who curate these data. The discovery and designation of minor planets begins with the accumulation of individual observations (e.g. right ascension, declination, time, observatory location, apparent magnitude) at the International Astronomical Union's Minor Planet Center (IAU MPC). These observations and their linkage to new or known minor planets are published by the MPC for independent analysis. Heliocentric orbits (ecliptic reference plane, reference epoch of J2000) defined by orbital elements semi-major axis $a$, eccentricity $e$, inclination $i$, argument of perihelion $\omega$, longitude of the ascending node $\Omega$, and mean anomaly $M$ for each minor planet are determined by fitting these observations (Section \ref{sec:orbit}). Following similar fitting processes, catalogs of orbital elements are curated at the MPC, by the Solar System Dynamics group at JPL, in Italy by a consortium that began at the University of Pisa, and by Lowell Observatory in Flagstaff, Arizona.

Lowell's \astorb catalog of asteroid orbits has grown organically into a modern relational database and associated web infrastructure.  However, the origins of \astorb trace back to the 1970's when Lowell astronomer Ted Bowell maintained a catalog of asteroid orbits on IBM punch cards which were input to an IBM 1130 computer\footnote{This computer had been installed at Lowell on 6 November 1966, seven years prior to Ted Bowell's arrival at the Observatory. From the Lowell Observatory annual report for 1966, the IBM 1130 computer was ``equipped with 8 K core memory, disk pack, card reader, paper tape reader, and line printer. [It] was installed in November 1966 in the soundproof, air-conditioned computer room of the Observatory's Planetary Research Center". This center along with the computer were funded by a NASA grant to facilitate studies related to the analysis of planetary imagery.}. These orbit cards were manually created by Bowell based on monthly circulars issued by the MPC. Orbits were also calculated for objects that were discovered ``in house" and then passed on to Brian Marsden, director of the MPC at the time. The software infrastructure used to process these orbits was developed by Lowell astronomer Lawrence Wasserman. 

A primary driver for maintaining this stack of punch card orbits was to have the ability to compute ephemerides for all known asteroids (a process described in Section \ref{sec:orbit}). These ephemeris predictions were used to find known asteroids on both recent and historic photographic plates, and later film exposures. At the time, many numbered asteroids were on the verge of being lost if new astrometry was not provided. Asteroids are generally numbered after observations have been obtained across four oppositions (though exceptions requiring fewer oppositions have been made for objects such as near-Earth asteroids). This was not the case for some objects. Work at Lowell focused on trying to solve this problem, which necessitated maintaining this local catalog of minor planet orbits. These early orbit computations were performed one object at a time and took about 5 minutes per object to complete on the IBM 1130. The full catalog contained fewer than 10,000 objects until 1980, and thus could be maintained in this manner with regular processing of the few hundred objects each month that needed to be updated.

Starting around 1977, use of the \astorb catalog was paired with the ongoing Palomar Planet Crossing Asteroid Survey (PCAS) led by Gene Shoemaker, Carolyn Shoemaker, and Eleanor Helin. PCAS was carried out at the Palomar 0.46m Schmidt with a primary objective of discovering near-Earth asteroids \citep{Helin79}. The PCAS films were developed soon after exposure and Bowell would get phone calls when objects with high rates of motion (consistent with a near Earth orbit) were discovered. Follow-up observations at Lowell would focus on confirmation and sometimes characterization including measurement of photometric phase curves, colors, and lightcurves. This process was quite efficient, such that orbits and physical characteristics were determined for new objects often before announcement of discovery had been made by the MPC.

In the mid 1980's the maintenance of this orbit catalog was made easier with the installation of a VAX 11/750 super minicomputer (about the size of a large washing machine). Integrations for individual objects would now only take a few seconds, and on board storage meant that the catalog could be saved to disk. This deprecated the use of punch cards and made \astorb a digital catalog, though it wouldn't be given that name until the next decade.

In the early 1990's Bowell and Finnish astronomer Karri Muinonen explored the mathematical problem of assessing orbit uncertainty and the associated error in positional (ephemeris) prediction \citep[e.g.][]{Muinonen93,Muinonen94}. This work led to the realization that an orbit database in the public domain that contained information beyond that published by the MPC would be of value to the research community \citep{Bowell93,Bowell94b}. By 1994, a flat ASCII file called \url{astorb.dat} containing orbits for 22,725 asteroids and updated daily was available for download from Lowell's anonymous \url{ftp} site\footnote{\url{astorb.dat} continues to be hosted and available for download at \url{https://ftp.lowell.edu/pub/elgb/astorb.dat}}. Though the format of this file has changed over the years, the basic information has remained constant and includes designations, absolute magnitudes, physical properties (slope parameter $G$, $B-V$ color, IRAS diameter and taxonomy), orbit details, and several parameters related to predicted ephemeris uncertainties.

Two years after \astorb became downloadable, the VizieR library of astronomical catalogs maintained at the Centre de Données astronomiques de Strasbourg (CDS) started hosting the ASCII catalog of orbits\footnote{ \href{https://cdsarc.cds.unistra.fr/viz-bin/cat/B/astorb}{https://cdsarc.cds.unistra.fr/viz-bin/cat/B/astorb}}. This was the first ever live catalog served by VizieR. It continues to be updated weekly and remains a valuable resource that sees $\sim500-1000$ requests per month through VizieR alone.

The combined VizieR and Lowell access points to the \astorb catalog helped to build a broad base of users. In the research community this is evidenced by regular citations dating back to the original public release \citep[e.g.][]{Morbidelli96}. In addition, the SkyBot minor planet identification tool \citep{Berthier06} was developed based on the orbits in \astorb and has since been leveraged by the Gaia mission \citep{Gaia16} to identify known asteroids in Gaia fields \citep{Carry21}. Commercial software packages that directly ingest the \astorb data file include \href{https://starrynight.com/}{Starry Night}, \href{https://www.bisque.com/product/theskyx-pro/}{The Sky}, and \href{https://skysafariastronomy.com}{Sky Safari}. This capability has helped \astorb maintain relevance to the hobbyist and amateur astronomy communities. As a whole, users of the \astorb system include professional and unpaid professional astronomers, educators and students, and general public who are scientifically curious. Given this user base, we have undertaken a major development effort to modernize the entirety of the \astorb system so that it is easier to maintain and will be able to accommodate ongoing growth of the minor planet catalog (Figure \ref{fig:firstobs}).

\section{Orbit Fitting and Integration} \label{sec:orbit}

Two key operations that support \astorb involve fitting orbits to sets of observations and the integration of orbits to various epochs. The basic methods involved in these process are well established \citep[e.g.][]{Gauss09}, but we describe them here to provide necessary context.

An object subject to only the gravitational force of the Sun will orbit the Sun in an ellipse which will not change with time. The ellipse can be defined by six parameters which are called the orbit's elements.  Two of the elements define the size and shape of the ellipse, three define how the ellipse is oriented in space, and one defines the location of the object in the orbit at a given time.

Unfortunately, the real world is not so simple as there are many other forces that change orbits over time.  These are: (1) The gravitational pulls of all the planets as well as all the other asteroids.  In practice, we account for all the planets, the Moon, which is important for Earth-approaching objects, and 16 of the most massive asteroids following JPL's DE440 planetary ephemeris \citep{Park21}. (2) General relativistic effects.  These are only significant for objects that come into the inner Solar System. (3) Non-gravitational forces due to cometary outgassing, which acts as a jet that can perturb an object in any direction. Our treatment follows the three-component formalism of \citet{Marsden73} in which accelerations in the radial, transverse and normal directions are treated independently. (4) The effect of radiation pressure forces, e.g. the Yarkovsky effect \citep{Bottke06}. The non-gravitational Yarkovsky terms used in our orbit solutions are retrieved from the \href{https://ssd.jpl.nasa.gov/tools/sbdb_lookup.html}{JPL Small Body Database.} (5) A second order term, $J_2$, that accounts for the gravitational potential of a non-spherical Earth, which is relevant to bodies that experience near-Earth encounters. The net result of these extra forces is that the orbital elements change with time so that we have to define a specific epoch associated with each set of elements for a given object.

If we have the six orbital elements at a given epoch (we  discuss how those are determined below), how do we convert them to a different epoch? First, note that the six orbital elements of a given object at a given epoch are mathematically equivalent to another set of six numbers -- the position (in Cartesian coordinates $x$, $y$, and $z$) and velocity ($\dot x$, $\dot y$, $\dot z$) of that object.  The proof of this is beyond the scope of this paper, but it is straightforward to convert elements to Cartesian coordinates/velocities and back again  \citep{Murray99}.  The forces mentioned above (gravity, relativity, non-gravitational) are all accelerations on the body in $x$, $y$, and $z$.  Given the current position, velocity, and acceleration, one may compute a new position and velocity a short time later (short enough so that the acceleration is effectively constant). Repeated application of this process over time allows going from an initial epoch to an arbitrary final epoch where one can then convert back to orbital elements at that final epoch. This process is called orbit integration. In practice, the process is not quite this simple. In a real calculation, the linear approximation just described is replaced by an eighth order integration which uses a nine point polynomial for the $x$, $y$, $z$ positions, velocities, and accelerations, with the polynomial recalculated at each time step. In addition, the time step size is variable and adjusted up or down as the net acceleration on the particle decreases or increases respectively. This method is essentially the same as that of \citet{Berry04}.

To determine the absolute magnitude $H$ for a given object we first take all observations with a reported magnitude and correct them to their equivalent magnitude in the Johnson $V$ filter using a standard list of $V-X$ values for every filter $X$ currently in the observations data base. These values are provided by the Minor Planet Center. We then compute $V$ for each observation of that object assuming H=0 and G=0.15.  This requires knowing the distance of the asteroid from the Sun, the distance of the asteroid from the Earth and the phase angle at the time of the observation. The average of the differences between our computed $V$ and the observed (and filter corrected) $V$ gives $H$.

The orbital elements and $H$ can be used to calculate an ephemeris. An ephemeris is a list as a function of time where a given object will appear in the sky (the object's right ascension and declination) and how bright it will be based on the observer's location.  We have to specify the location because parallax will affect the apparent position of the object on the sky.  Since the elements change with time, the ephemeris has to be generated from integrated elements.

But, we still need to explain where the orbital elements come from in the first place. Given a set of observations of a body over time (from one or more observatories), we can do a least squares fit to determine the six orbital elements\footnote{For our purposes, observations are sourced from the \href{https://minorplanetcenter.org/iau/ECS/MPCAT-OBS/MPCAT-OBS.html}{MPC observations files} and do not include Doppler delay radar measurements.}. A starting set of elements is integrated to each of the observation dates and an ephemeris is calculated for that date/location.  We minimize the observed minus calculated right ascension/declination coordinates in a least squares sense until we find the elements which best fit all the observations in the data set. In an automatic fit, outliers are rejected until all have residuals below 2.3 arcseconds (an empirical value used for many years). For a batch of 100,000 asteroids, a small number, typically $\sim$20-50, fail to fit automatically. In those cases, a manual fit with a larger rejection threshold (3, 4, or 5 arseconds) is performed. Weighting individual observations in these fits is difficult due to the wide diversity of data. Data for some asteroids go back as far as 1801 and were obtained by many different techniques (e.g. visual, photographic, CCD). In addition, the precision of astrometric reference stars has improved over the years, with star positions from the Gaia survey \citep{Gaia22} currently providing the best available reference catalog. As such we do not weight observations in our orbit fits.

Note that, in practice, when an object is first discovered, it may have a very short observational arc and, as a result, its elements determined by the least squares procedure may have large uncertainties, which will result in large errors in its ephemeris when the elements are integrated to times well separated from the discovery epoch.  Each time the object is re-observed, the resulting set of orbital elements and its ephemeris become better constrained.  The Minor Planet Center will assign a permanent number to an asteroid once its orbit becomes accurate enough that the error in its ephemeris will always be small.  On the other hand, if a newly-discovered object is not re-observed, its error can become large enough that it becomes effectively lost.  

A simple assessment of the quality of orbits in \astorb can be achieved by comparing our orbital elements to those from JPL and the MPC. From an observational perspective, the ultimate test of ``which orbit solution is best?" would involve on-sky testing of measured positions versus ephemeris predictions, a full study of which is beyond the scope of this work. Empirically, we have found that all three systems produce similar ephemeris predictions and that any differences are most pronounced for short arc objects.

\begin{figure}
     \centering
     \begin{subfigure}[b]{0.9\textwidth}
         \centering
         \includegraphics[width=\textwidth]{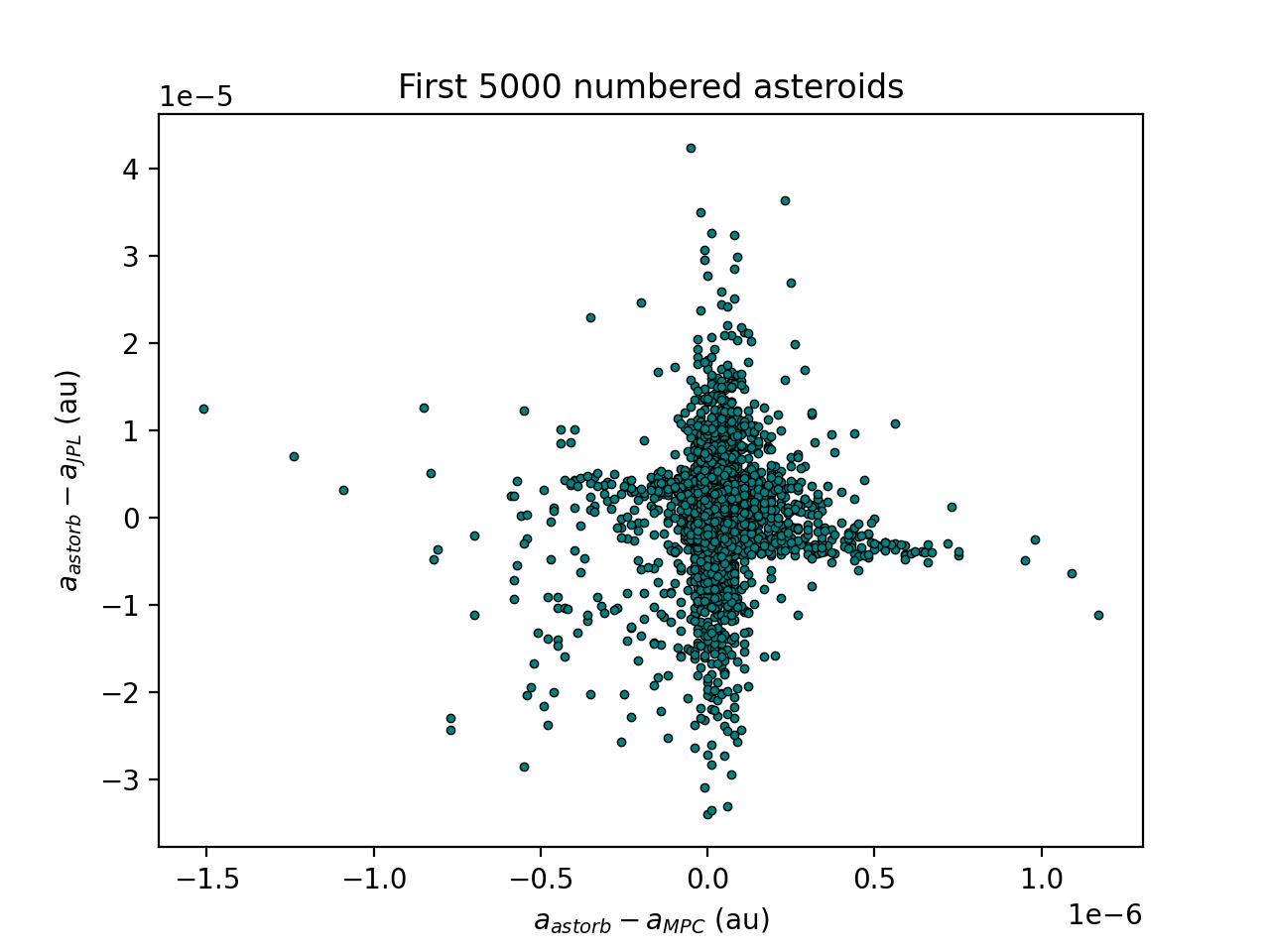}
     \end{subfigure}
     \begin{subfigure}[b]{0.9\textwidth}
         \centering
         \includegraphics[width=\textwidth]{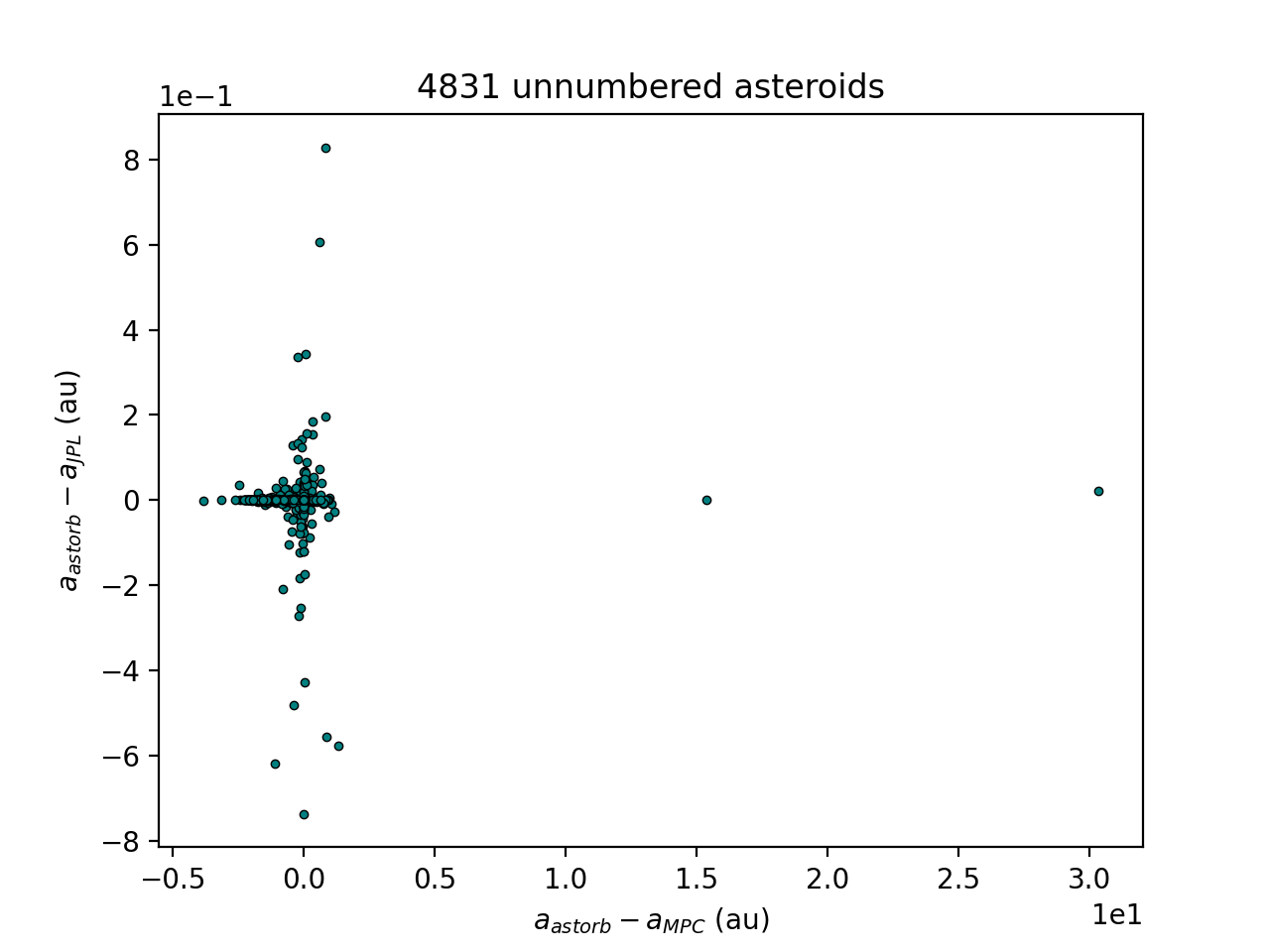}
     \end{subfigure}
    \caption{Comparison of semi-major axes $a$ for the first 5000 numbered asteroids and a set of recently discovered unnumbered asteroids. Axes are the differences in semi-major axis for \astorb and MPC versus \astorb and JPL. Unsurprisingly the orbits for the numbered objects are in good agreement, whereas unnumbered objects with short arcs can have orbits that different across the three systems.}
    \label{fig:orbcomp}
\end{figure}

We consider two samples of objects in an \astorb-JPL-MPC comparison: the first 5000 numbered asteroids and a similar number (4831) of recently discovered objects with arc lengths as small as 1 day. Figure \ref{fig:orbcomp} shows the comparison of semi-major axes for these two samples and Table \ref{tab:orbcomp} provides statistics. 

Unsurprisingly the elements for the numbered objects are very similar in all catalogs with mean offsets typically at the $10^{-5}-10^{-8}$ level, depending on the orbital element. The standard deviations of these offsets are similarly small, showing that there are no large systematic biases in the numbered asteroid sample. In general the MPC and \astorb elements are in closer agreement for numbered asteroids than JPL and \astorb. It is unclear why that would be the case.

There are much larger offsets in the unnumbered sample. However, the origin of these offsets is difficult to fully understand. For the particular version of the orbit catalogs considered here, the largest offsets (e.g. semi-major axis differences $>5$ au or inclination differences $>30^\circ$) are seen in the \astorb to MPC comparison. For example, at the time of writing asteroid 2022 QY7 had a semi-major axis offset of 30.2 au in the \astorb and MPC catalogs, but JPL and \astorb agreed to within 0.02 au. Orbit offsets are expected for objects like 2022 QY7, which had a short observational arc of only 2 days on a 190 year orbit period. Furthermore, the number of individual observations accepted for the orbit fits was not the same: the \astorb orbit was based on 7 observations while the MPC orbit was based on 13. However, \astorb-MPC offsets are systematically larger than \astorb-JPL offsets for such short arc, long period objects. This bias can been seen in the mean and standard deviations of the element comparison, where the \astorb-MPC values are $\sim2$ orders of magnitude larger than the JPL to \astorb numbers (Table \ref{tab:orbcomp}). Following a correspondence with the MPC, these differences were investigated and fixed with an update to the MPC catalog. As such, these large offsets may have simply been a consequence of differing update cadences to the orbit catalogs. To facilitate future comparisons of this nature, the MPC is in the process of constructing a dedicated web service to compare MPC orbits with those computed at JPL and at Lowell. The results will be regularly updated and published.

\begin{table}[h!]
\begin{tabular}{ lll } 
\hline
Element difference      & Mean $\pm~\sigma$ , Numbered  & Mean $\pm~\sigma$, Unnumbered \\ 
\hline
\hline
$a_{astorb} - a_{JPL}$  & 3.8e-6 $\pm$ 5.7e-6 au    & 0.003 $\pm$ 0.03 au \\
$a_{astorb} - a_{MPC}$  & 6.5e-8 $\pm$ 1.2e-7 au    & 0.14 $\pm$ 0.6 au \\
$e_{astorb} - e_{JPL}$  & 1.0e-6 $\pm$ 1.6e-6       & 0.0007 $\pm$ 0.005 \\
$e_{astorb} - e_{MPC}$  & 6.8e-8 $\pm$ 1.1e-7       & 0.03 $\pm$ 0.1 \\
$i_{astorb} - i_{JPL}$  & 1.4e-5 $\pm$ 2.1e-5 deg   & 0.01 $\pm$ 0.1 deg \\
$i_{astorb} - i_{MPC}$  & 1.2e-5 $\pm$ 1.5e-5 deg   & 1.2 $\pm$ 4.3 deg \\
\hline
\end{tabular}
\caption{Statistics (mean, standard deviation) on the differences in orbital elements $a$, $e$, and $i$ in the \astorb, JPL and MPC catalogs. These are calculated for samples of 5000 numbered and 4831 unnumbered asteroids. Means are calculated on the absolute values of the differences.}
\label{tab:orbcomp}
\end{table}

These processes of integration and orbit fitting have remained relatively unchanged over the lifetime of \astorb with some current operational subroutines dating back to the 1970's. As we look ahead to challenges associated with growth of the minor planet catalog, we are revisiting these core functions. Adoption of the hybrid symplectic, GPU-based orbit integrator GENGA \citep[Gravitational Encounters in N-body simulations with GPU Acceleration,][]{Grimm14} will be foremost in those efforts and will be the topic of a future publication. Preliminary tests with GENGA suggest at least a $30\times$ boost in integration speed relative to our current CPU-based, direct integration scheme. General guiding principles when implementing such upgrades to \astorb are to maintain existing functionality for end users, and to provide resources to facilitate telescopic observations. Generally, we do not know in detail for every object how our orbit solutions compare to those at the MPC and JPL, but we meet our objectives if the orbits and ephemerides are well enough defined that an observer can successfully find their target (given reasonable estimates on ephemeris errors) with a telescope.

\section{Maintenance and Infrastructure} \label{sec:backend}

\subsection{Maintaining \astorb} \label{subsec:maintain}

Since its inception \astorb has relied on observations and designations provided by the MPC. As new observations become available they are fit with standard methods (Section \ref{sec:orbit}) to update our catalog of heliocentric orbits. The observations are downloaded from the MPC on a semi-monthly cadence. Orbits for objects announced by Minor Planet Electronic Circulars (MPECs) are added outside of this cadence. As soon as an MPEC is issued, the orbit from the MPC is added to \astorb and serves as a placeholder until we fit the observations ourselves at the time of the semi-monthly update. Parameters specific to \astorb, such as the covariance associated with our orbit fit and ephemeris-related parameters (Section \ref{subsec:db}), follow this update cadence.

When an orbit is created in \astorb it is integrated (Section \ref{sec:orbit}) to a set of epochs at pre-defined 100 day intervals. There are currently 103 of these 100-day epochs running from 1997-12-17 to 2025-11-20. Orbits are stored at each of these epochs to facilitate calculation of positions outside of these 100-day intervals, ie the longest any object would have to be integrated is 50 days from the nearest epoch of stored elements. Our system can calculate positions outside of this date range, but integrations will take longer for dates much earlier than 1997 or later than 2025. Future work will involve the expansion of 100-day epochs to dates later than 2025.

At present the fitting and integrating of orbits for the semi-monthly update takes nearly a day of computing time on 20 CPUs to process the tens or even hundreds of thousands of objects with new observations in any given month. Thanks to improvements in computing power this mode of operation has remained viable for many years, however this approach will be strained when LSST increases discovery rates by a factor of $\sim10$ \citep{Jones09}. Given this concern, our modernization efforts have focused on upgrading the full hardware and software infrastructure that maintains \astorb. 

\subsection{Hardware} \label{subsec:hardware}

The hardware infrastructure to support \astorb consists of a database server, compute server, and a web server to host \url{asteroid.lowell.edu}. We provide a brief overview here of the server hardware as it stands in mid 2022. All three servers communicate over a 10 Gbit backbone, and each has been built to optimize performance for particular tasks. The operating systems for all servers are current LTS (long term support) Ubuntu distributions.

The database (Section \ref{subsec:db}) is hosted on a 1U SuperMicro rack server. The server has a 24-core AMD EPYC 7402P processor, 128GB of DDR4 ECC memory, and 8 x 2TB NVMe (non-volatile memory express) SSDs (solid state drives) setup in a RAID-Z configuration. Two additional 256GB NVMe SSD drives act as redundant system boot disks. The primary design requirement for this server was optimization of read-write performance to the database.

Our compute server was designed to optimize both CPU and GPU computing tasks that are essential to \astorb maintenance. It is a 4U SuperMicro rack powered by two 20-thread Intel Xeon Silver 4210 processors. It hosts 128GB of DDR4 ECC memory, and a 1TB NVMe SSD. To support our work in GPU-based orbit integration (Section \ref{sec:orbit}), this sever is also equipped with a NVIDIA RTX 2080Ti GPU.

The web server is a virtual machine managed by VMware, and is configured with eight 2.2 GHz Intel Xeon CPUs and 8GB of RAM memory.

\subsection{Software} \label{subsec:software}

Prior to 2015, the software to maintain \astorb was primarily written in Fortran, the database existed as Fortran binary files, and the website and associated online tools were supported with Ada, Fortran, and IDL. Concerted effort has been made to modernize this software foundation. A general philosophy in development has been to adopt open source and actively developed tools whenever possible.

We have adopted \href{https://www.python.org/}{Python} as the fundamental programming language that supports much of \astorb. Many of the legacy tools written in Fortran have been converted to Python. Fortran is well developed and performant and shines in specialized cases. However, Python affords more flexibility, is modern, and highly extensible, making it a powerful choice for development. Furthermore, Python is  widely implemented in the scientific community, with \href{https://www.astropy.org/}{The Astropy Project} \citep{Astropy22} and \href{https://sbpy.org}{sbpy} \citep{Mommert19} being notable examples. 

Many backend procedures and scripts are written in or being converted to Python and \href{https://asteroid.lowell.edu}{asteroid.lowell.edu} is built using \href{https://flask.palletsprojects.com/en/2.1.x/}{Flask}, a mature and fast Python web framework. Our PostgreSQL database on which our tools fundamentally rely (Section \ref{subsec:db}), is managed using the Object Relational Mapper \href{https://www.sqlalchemy.org/}{SQLalchemy}, a Python package for abstracting raw SQL. By using these tools, we hope to make future development fast and approachable. The data underlying astorbDB (Section \ref{subsec:db}) are made available through a \href{https://graphql.org/}{GraphQL} application programming interface (API) at \url{https://astorbdb.lowell.edu/v1/graphql} (see Section \ref{sec:api} for more details). 

The codebase for \href{https://asteroid.lowell.edu}{asteroid.lowell.edu} and the backend scripts for managing and running astorbDB are version controlled with \href{https://git-scm.com/}{Git} with repositories becoming public as they mature. \href{https://asteroid.lowell.edu}{asteroid.lowell.edu} is containerized using \href{https://www.docker.com/}{Docker} to make deployment simple and predictable. We follow the Continuous Integration and Deployment model of code release, making use of features and pipelines provided by \href{https://gitlab.com}{GitLab}. This model and these tools allow our small team to continuously and reliably develop the codebase.

\subsection{Database} \label{subsec:db}
  
\begin{figure}[h!]
    \includegraphics[width=\textwidth]{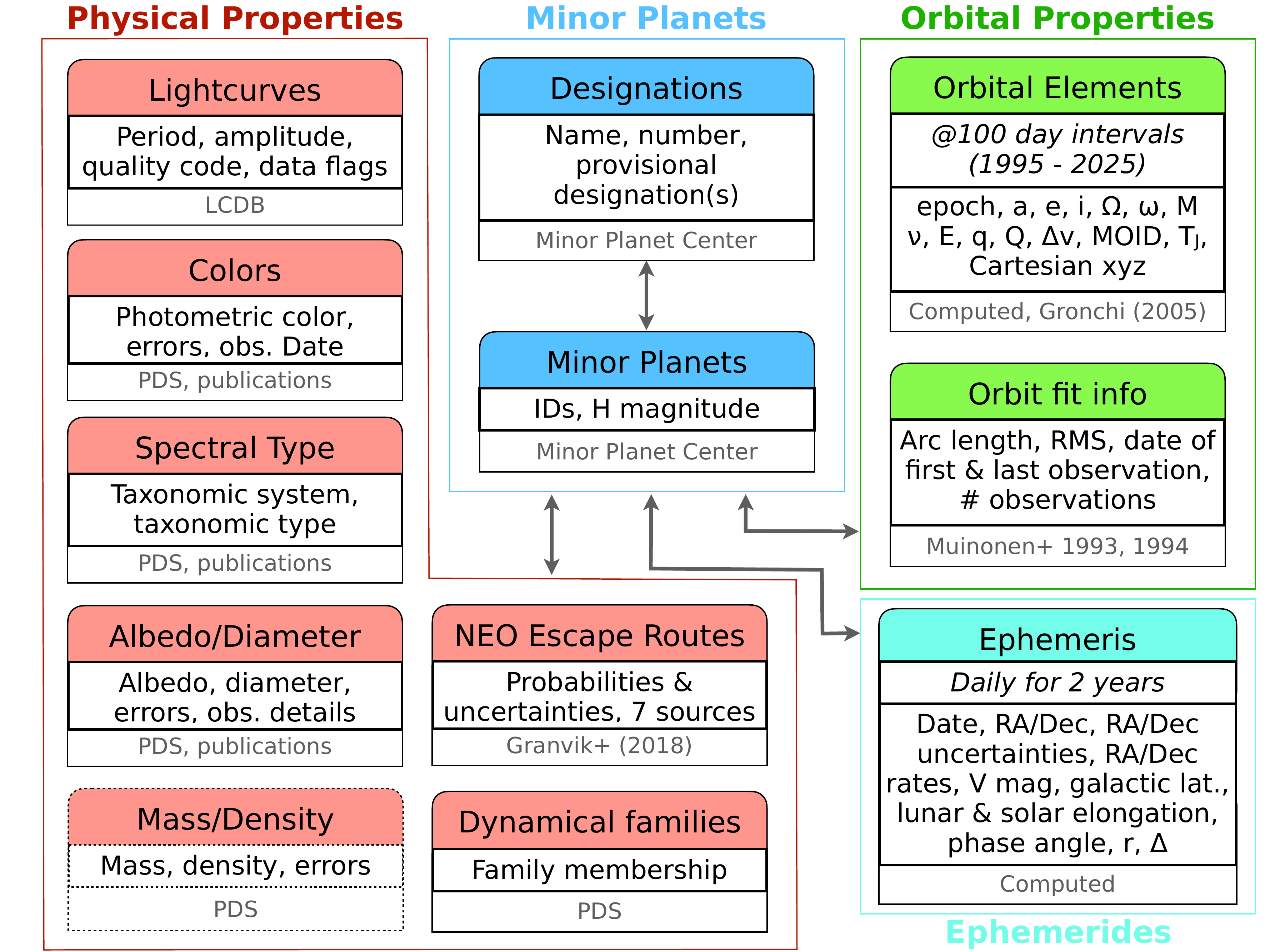}
    \caption{Cartoon representation of our database schema. Citations for individual physical properties are given in Table \ref{tab:phys}. These data generally come from a combination of the NASA Planetary Data System (PDS) and select publications. Computations related to orbits and ephemerides are based on \citet{Gronchi05}, \citet{Muinonen93}, \citet{Muinonen94}, and the processes described in Section \ref{sec:orbit}. Physical properties associated with mass and density are not currently in the database, but will be added in future updates.}
    \label{fig:schema}
\end{figure}

A major part of our modernization effort has been the migration of \astorb to a \href{https://www.postgresql.org}{PostgreSQL} relational database, which we refer to as astorbDB. This framework is an important upgrade as the size of the minor planet catalog grows and complexity is added to \astorb. A cartoon of the astorbDB schema is shown in Figure \ref{fig:schema}. This schema can broadly be divided into four main categories: minor planets, orbital properties, ephemerides, and physical properties.

At the core of astorbDB are the minor planet and designation tables. These track the designations (numbers, names, provisional, and primary) and absolute magnitude ($H$) of every object. The designations are ingested and maintained directly from \href{https://minorplanetcenter.net/data}{data files} published by the MPC. Every object must have a primary designation and an $H$ magnitude. For the vast majority of objects, $H$ is derived from the orbit fitting procedure (Section \ref{sec:orbit}) and follows the $HG$ formalism developed by \citet{Bowell89}. However, there are currently about 1000 objects without $H$ at the MPC. For these we assign a value of $H = 17$. Historically this coincided with the most common $H$ found by ongoing discovery surveys. Over time surveys have become more effective, so this value is infrequently adjusted. It used to be 15, future updates may see this adjusted to a value of 19 or 20. Regardless of the value, these objects are all unnumbered and generally have short arcs of observations that span just a few nights, meaning their orbits are poorly constrained\footnote{There are a few dozen interesting exceptions of objects that were discovered primarily around 2010 by the WISE satellite \citep{Mainzer11}. These objects can have observational arcs of many tens or a few hundred days, but don't have accompanying visible wavelength observations from which an $H$ magnitude could be derived. These objects are typically small (few km) and/or low albedo Main Belt asteroids and will need to be recovered by future surveys.}. As such, any attempts to obtain follow-up observations of these objects would be hampered by large ephemeris uncertainties. In other words, most of these objects are effectively lost, thus assigning an arbitrary $H$ has little impact on the overall functionality of the \astorb system. However, the somewhat arbitrary assignment of $H=17$ could influence population studies interested, for example, in the number of objects as a function  of sizes. This issue will need to be considered in any future updates to this set of objects without $H$ values.

Associated with each minor planet are orbit tables. These include the 103 epochs of orbital elements spanning the years 1997 - 2025, and details of the information used to generate each orbit. We store ``asteroidal" orbital elements, namely semi-major axis $a$, eccentricity $e$, inclination $i$, argument of perihelion $\omega$, longitude of the ascending node $\Omega$, and mean anomaly $M$ at each epoch. To facilitate internal calculations, namely conversion of Keplerian elements to Cartesian coordinates, we also store true anomaly $\nu$, eccentric anomaly $E$, perihelion distance $q$, and aphelion distance $Q$. Several computed quantities derived from orbital elements are also stored (Section \ref{sec:dyn}). 

Quantities associated with the observations and orbit fitting are stored in a separate table. These include the arc length spanned by the observations, the RMS of the orbit fit, the dates of first and last observation, the numbers of observations and corresponding apparitions, and the covariance matrix associated with the orbit fit. Other computed metrics based on \citet{Muinonen93} and \citet{Muinonen94} include an orbit quality parameter (analogous to the MPC's uncertainty parameter), the current ephemeris error (CEU), the rate of change in the CEU, the date of the CEU, and information about the next peak ephemeris uncertainty (PEU). The latter represents the next maximum in the ephemeris uncertainty and usually indicates an optimal time to make astrometric observations for orbit improvement. The date of the next PEU (i.e. the local maximum) is calculated, as are the date of the maximum PEU over the next 10 years and the date of the greatest PEU in the next 10 years if two hypothetical observations of arcsecond precision are made on the date of the next PEU.

The largest single table in our schema is the ephemeris table where pre-computed daily ephemerides are generated for every object from the present day up through 2 years in the future. This table is updated every day with an extra day appended to the end of the table while yesterday is deleted. This results in nearly 1 billion rows (1 million objects $\times$ 720 days) with right ascension, declination, non-sidereal rates, positional uncertainties, galactic latitude, lunar elongation, solar phase angle, heliocentric distance, geocentric distance, solar elongation, and predicted V magnitude available for every date. The primary function of this ephemeris table is to facilitate queries on the observability of targets on a given date or range of dates. Using the table partitioning function introduced in PostgreSQL 10, we have partitioned the ephemeris table by date. With our current infrastructure, this results in $\sim1$ second queries to determine which specific members of a given population, e.g.~near-Earth objects, are observable on any given night in the next two years.

The final category of tables are those related to physical properties. This is a recent addition to \astorb and continues to grow as we ingest new data from a variety of sources. We generally refer to these data as {\it survey data products} since many are the result of focused surveys, for example photometric colors of asteroids from the Sloan Digital Sky Survey \citep{Ivezic10}. However, we use this term loosely as some of these survey tables contain compiled data sets (e.g. data products compiled from multiple sources) or are the result of numerical models. We have designed the survey table infrastructure to be sufficiently flexible to handle this diversity. For every set of survey data that is ingested, we assign a survey type and measurement technique(s). Currently supported survey types include albedo, photometric colors, dynamical family, lightcurves, spectral taxonomy, and NEO escape routes (as defined by \citet{Granvik18}). These types, techniques, and associated data are described in greater detail in Section \ref{sec:phys}. Future additions to the survey types will include mass and density, shape information (e.g. axis ratios), and information on the presence of satellites/binarity. 

For all data that are stored in \astorb, from both internal and external sources, we maintain extensive citation capability. This includes information such as URLs to websites, bibcodes to publications, or names of individuals responsible for performing a set of calculations. Whenever possible we provide these citations to the end user so that those responsible for the original data are properly credited.

\section{Dynamical calculations} \label{sec:dyn}

Upon creation of an object in \astorb and integration of its orbit to each of the 100-day epochs (Section \ref{sec:orbit}), several dynamical calculations  are performed. These calculations are performed for every 100-day epoch.

The first of these calculations determines dynamical type based on analytical criteria. A scheme to define dynamical types was developed with requirements that every object in the catalog be assigned at least one dynamical type, and that any given object not be assigned logically exclusive types. For example, an object can not be both a Main Belt asteroid and a Jovian Trojan. Analytic expressions were defined based on orbital parameters semi-major axis ($a$), eccentricity ($e$), and inclination ($i$). These definitions (Table \ref{tab:dyntype}) were based as much as possible on precedent in the literature \citep[e.g.][]{Kirkwood82,Zellner85,Jewitt05,Lykawka07,Gladman08}. Eight super types -- Near-Earth Object (NEO), Mars Crosser (MC), Main Belt Asteroid (MBA), Jovian Trojan, Jupiter Crosser, Damocloid, Centaur, and Trans Neptunian Object (TNO) -- encompass the entire catalog. Several of these super types also contain sub-types. While such definitions have historically been assigned to individual populations, it is less common to see a holistic approach to classifying all asteroids in the Solar System. The SkyBot service \citep{Berthier06} adopts a similar approach to us with slightly different definitions for some of the sub-types.

\begin{table}[h!]
\scriptsize
\begin{tabular}{ |p{0.4\linewidth}|p{0.45\linewidth}|p{0.3\linewidth}|p{0.05\linewidth}| } 
 \hline
 \bf Dynamical Type & \bf Analytic Definition & \bf Note & \bf Ref. \\ 
 \hline
 \hline
 {\bf Near-Earth Object (NEO)} & $q < 1.3$ & - & [1] \\ 
 NEO: Apollo & $a \ge 1.0$, $q \le 1.017$ & - & [1] \\ 
 NEO: Aten & $a < 1.0$, $Q > 0.983$ & - & [1] \\ 
 NEO: Amor & $a > 1.0$, $1.017 < q < 1.3$ & - & [1] \\ 
 NEO: Atira & $a < 1.0$, $Q < 0.983$ & - & [1] \\ 
 NEO: Potentially Hazardous Asteroid (PHA) & $H \le 22$, $MOID \le 0.05$ & - & [1] \\ 
 {\bf Mars Crosser (MC)} & $1.3 \le q \le 1.666$ & An object whose orbit crosses that of Mars and is not an NEO & - \\ 
 {\bf Main Belt Asteroid (MBA)} & $q > 1.666$, $a < 4.8$ & Bounded by MCs on the inner edge and Jovian Trojans on the outer & - \\ 
 MBA: Inner belt & $2.0 \le a \le 2.5$ & Defined by Kirkwood gaps & [2] \\ 
 MBA: Middle belt & $2.5 < a \le 2.82$ & Defined by Kirkwood gaps & [2] \\ 
 MBA: Outer belt & $2.82 < a < 4.8$ & Defined by Kirkwood gaps & [2] \\ 
 MBA: Hildas & $3.7 \le a \le 4.2$, $e \le 0.3$ & - & [3] \\ 
 {\bf Jovian Trojan} & $4.8 \le a \le 5.5$, $e \le 0.3$ & Criteria encompass all Trojans defined by [4] & [4] \\ 
 {\bf Jupiter Crossers (JC)} & $4.8 \le a \le 5.5$, $e > 0.3$ & High eccentricity objects on Jupiter crossing orbits & - \\ 
 {\bf Damocloid} & $\frac{5.204267}{a} + 2 cos(i)\sqrt{\frac{a}{5.204267}(1-e^2)} < 2$ & - & [5] \\
 {\bf Centaur} & $ 5.5 < a \le 30.0709$ & Bounded by Jovian Trojans on the inner edge and Neptune on the outer & - \\
 {\bf Trans Neptunian Object (TNO)} & $ a > 30.0709$ & Outside of Neptune's orbit & - \\
 TNO: Cold Classical & $i < 5$,  $q \ge 37, 37 \le a \le 40$ & - & [6] \\
   & ~~~OR &  &  \\
 & $i < 5$, $q \ge 38, 42 \le a \le 48$ &  &  \\
TNO: Hot Classical & $i > 5$, $q \ge 37$, $37 \le a \le 48$ & - & [6] \\
 TNO: Scattered Disk Object & $e > 0.4$, $25 \le q \le 35$ & - & [7] \\
 TNO: Detached & $e > 0.25$, $q > 40$ & - & [7] \\
\hline
\end{tabular}
\caption{Dynamical type definitions based on absolute magnitude ($H$), MOID, semi-major axis ($a$), eccentricity ($e$), inclination ($i$), perihelion ($q$), and aphelion ($Q$). All distances are in units of AU, all angles in degrees. Entries in bold are primary types that may contain multiple sub-types. References are [1] JPL Center for Near Earth Object Studies (\url{https://cneos.jpl.nasa.gov}), [2] \citet{Kirkwood82}, [3] \citet{Zellner85}, [4] IAU Minor Planet Center list of Jupiter Trojans (\url{https://minorplanetcenter.net//iau/lists/JupiterTrojans.html}), [5] \citet{Jewitt05}, [6] \citet{Lykawka07}, [7] \citet{Gladman08}.}
\label{tab:dyntype}
\end{table}

This scheme for dynamical type assignment generally produces results on individual objects that are fully consistent with classifications that appear in the literature. However, this scheme also facilitates `needle in the haystack' searches, i.e. the identification of unusual objects. For example, asteroids 514107 Ka'epaoka'awela, 2007 VW266, and 2016 YB13 stand out as (currently) the only three Jupiter Crossers with retrograde (inclination $>90^\circ$) orbits. Such exotic objects can help probe planet formation and evolution processes \citep[e.g.][]{Connors18,Morbidelli20}. We also find objects that span multiple dynamical types. As examples, asteroid 2019 EJ3 is classified as an Amor NEO, a Damocloid, and a TNO, while asteroid 1999 XS35 is currently the only known example of an Apollo PHA, Damocloid, and Centaur. These objects offer surprising examples of multiple classifications that raise interesting questions for the community, outside the scope of our work, related to whether it makes sense for such varied classifications (e.g. TNO and NEA) to be allowed for individual objects.

The second dynamical computation involves the minimum orbit intersection distances (MOID) relative to each of the 8 major planets. MOID provides an estimate of the minimum distance between two orbits and is used in the definition of asteroids that may pose an impact hazard to Earth \citep{Bowell94}. We have adopted the FORTRAN90 critical points program by \citet{Gronchi05} with adaptations implemented to handle standard input/output and to better interface with our code base. MOID is computed for all 8 major planets -- Mercury, Venus, Earth, Mars, Jupiter, Saturn, Uranus, Neptune -- resulting in cataloging of nearly 1 billion individual values ($\sim1$ million asteroids $\times 103$ epochs $\times$ 8 major planets). We use Earth MOID values computed for NEOs to classify potentially hazardous asteroids (Table \ref{tab:dyntype}).

The \citet{Gronchi05} code is configured to attempt 10 numerical iterations to reach convergence on the MOID for any given pair of orbits. In $\gg$99\% of cases this results in convergence. Convergence tends to fail most often for objects with high eccentricity. At any given epoch only about 10 objects fail to produce a full suite of 8 MOID values and more than 90\% of these have eccentricity $>0.95$. The most common MOID that fails to compute is with the planet Mercury.

Following \citet{Shoemaker78} we also compute $\Delta v$ for all near-Earth objects (NEOs), which is a metric for the energy required to transfer a spacecraft trajectory from low-Earth orbit to rendezvous with a given object. Low $\Delta v$ objects are of high interest for spacecraft exploration and activities related to in space resource utilization \citep[e.g.][]{Elvis11}. Other data systems provide more detailed assessment of NEO accessibility, for example \href{http://www.ecocel-database.com}{ECOCEL} (Explotation des Ressources des Corps Celestes) and the NASA/JPL \href{https://cneos.jpl.nasa.gov/nhats/}{NHATS} (Near-Earth Object Human Space Flight Accessible Targets Study) list.

Finally, we compute for all bodies the Tisserand invariant relative to Jupiter $T_J$, a parameter commonly used to distinguish Main Belt asteroids from Jupiter-family comets.

\section{Physical properties} \label{sec:phys}

While \astorb has traditionally been a catalog of orbital elements, we began in 2018 an expansion to add derived physical properties to the database. The motivation of this expansion was not to replicate the function of public archives such as NASA Planetary Data System (PDS) where original data (e.g.~fits files from telescopic surveys) are stored, but instead to provide easy access to a compilation of derived physical properties. As an example of this distinction: we focus on ingesting derived albedo, but do not accommodate the original flux measurements or images used to determine that property. This compilation spans a wide range of physical properties (Table \ref{tab:phys}) and is actively curated. Some data sets, like the asteroid lightcurve database \citep[LCDB,][]{Warner09}, are automatically ingested on a regular cadence, others are ingested only when updates become available, and some represent a single one-time ingest.

These physical data are sourced from the PDS, publications, and  project websites. The details of these sources are discussed in the following sub-sections related to each category of physical property. This compilation is far from complete as the number of relevant publications is large and often, particularly for historical literature, contains data in formats that are not easily ingested. Generally, we have focused on publications that are the outcome of surveys across many objects and not those that offer detailed information about individual objects. Inevitably this process is imperfect and many publications will have been missed. As the \astorb compilation of physical properties continues to grow, we welcome investigators to contact us with new data sets that could be made accessible in our system.

Our database schema includes assignment of measurement technique(s) with each physical property. These techniques include astrometry, visible photometry, near-IR photometry, mid-IR photometry, multiple filter set photometry, Doppler delay radar imaging, visible spectroscopy, near-IR spectroscopy, spacecraft or in situ data, direct imaging and/or adaptive optics,  occultation, simulation, polarimetry, and literature compilation across multiple measurement techniques. The intent in storing this information is to provide the end user information to help compare properties derived in different ways. For example, various asteroid taxonomic systems have be developed over the years  \citep[e.g.][]{Tholen89,Bus02,Demeo09}. Some of those are based on broad band photometry at visible wavelengths \citep[e.g.][]{Carvano10}, some based on visible wavelength spectroscopy \citep[e.g.][]{Bus02}, are some are based on combinations of properties such as spectro-photometry, albedo, and polarimetry \citep[e.g.][]{Tholen84,Bowell78}. With access to these measurement methods, results can be sorted or prioritized based on a given investigator's preference.

\begin{table}[h!]
\scriptsize
\begin{tabular}{ |p{0.25\linewidth}|p{0.5\linewidth}|p{0.25\linewidth}| } 
 \hline
 \bf Physical Property & \bf Data products & \bf Data source(s) \\ 
 \hline
 \hline
  Rotational lightcurves & Period, amplitude(s), quality code, boolean flags (ambiguous period, non-principal axis rotator, sparse data, data from wide field survey) & [1] \\
   \hline
Albedo, diameter & Albedo, diameter, NEATM $\eta$ (when applicable), upper and lower error bars, number of observations and number of band passes used for derivations & [2,3,4,5,6,7,8,9,10,11] \\
   \hline
Photometric colors & Colors from surveys in filter sets SDSS $ugriz$, Jonhson-Cousins UBVR$_c$I$_c$, and 2MASS JHK; associated error bars; JD of observation & [12,13,14,15,16,17,18] \\
 \hline
  Taxonomy & Taxonomic type, taxonomic system & [19,20,21,22,23,24] \\
 \hline
  Dynamical family & Membership in family with identified parent body & [25] \\
 \hline
  NEO escape routes & Probabilities and uncertainties for seven NEO escape routes & [26] \\
\hline
\end{tabular}
\caption{Compiled physical properties available in \astorb tables. Data sources are: [1] \citet{Warner09}, [2] \citet{Tedesco04}, [3] \citet{Shevchenko07}, [4] \citet{Delbo06}, [5] \citet{Usui11}, [6] \citet{Mainzer19}, [7] \citet{Lupishko14}, [8] \citet{Morrison07}, [9] \citet{Trilling10}, [10] \citet{Trilling16}, [11] \citet{Trilling16b}, [12] \citet{Tedesco05}, [13] \citet{Sykes10}, [14] \citet{Chatelain16}, [15] \citet{Baudrand07}, [16] \citet{Ivezic10}, [17] \citet{Neese14}, [18] \citet{Fornasier08}, [19] \citet{Binzel19}, [20] \citet{Demeo13}, [21] \citet{Fulchignoni00}, [22] \citet{Perna18}, [23] \citet{Neese10}, [24] \citet{Hasselmann11}, [25] \citet{Nesvorny15}, [26] \citet{Granvik18}.}
\label{tab:phys}
\end{table}

\subsection{Rotational lightcurves} \label{subsec:lc}

By far the most widely used compilation of lightcurve properties is the asteroid Lightcurve Database \citep[LCDB,][]{Warner09}. This database is updated roughly every month and represents a compilation of lightcurve properties transcribed from relevant publications. Indication of a preferred solution (e.g. period and amplitude) for a given object and a quality code for each published lightcurve are two of the unique aspects of the LCDB. The LCDB contains a variety of ancillary data such as albedo and taxonomic class that we do not ingest as we pull those data from primary sources. The LCDB data products that we do ingest include the preferred lightcurve period and amplitude as determined by the LCDB curators, the lightcurve quality code \citep{Harris79}, and four Boolean flags to indicate whether the body is in non-principal axis rotation, the solution is ambiguous, the solution came from sparse data, and the data were obtained as part of a wide-field survey. Our system is configured to ingest updates from the LCDB soon after they become available. Given the nature of the LCDB data we determined that it was too difficult to automatically parse individual citations for each object. As such we provide an overall citation to the database \citep{Warner09} and encourage \astorb users to find object specific citations on the \href{https://minplanobs.org/MPInfo/php/lcdb.php}{LCDB website}.

\subsection{Albedo and diameter} \label{subsec:albedo}

Since the 1970's, numerous surveys operating at mid-infrared wavelengths ($\sim3.5-30~\mu m$) have measured thermal emission from minor planets. These thermal flux measurements are generally fit with models in which diameter and albedo are simultaneously solved. One such widely used model is the near-Earth asteroid thermal model \citep[NEATM,][]{Harris98}, which includes a free parameter known as the thermal beaming parameter $\eta$. This parameter compensates for a lack of detailed knowledge of thermophysical properties such as thermal inertia, but can provide coarse insight on some surface characteristics \citep[e.g.][]{Delbo03,Moskovitz17}. The albedo table in \astorb has been designed to accommodate these properties (albedo, diameter, and NEATM $\eta$ when applicable) along with their associated upper and lower error bars. In addition, we include the number of photometric band passes and the number of individual observations in which the original thermal data were obtained. For example, during the cryogenic phase of the WISE mission (January-September 2010), four band passes at 3.4, 4.6, 12 and 22 $\mu m$ were employed, whereas just the two short-wavelength bands were used in the post-cryogenic NEOWISE mission \citep{Mainzer11}. These numbers can help users decide which albedo-diameter values to adopt for their own science. For example, asteroid (1766) Slipher was observed in the thermal-IR as part of asteroid surveys conducted by the AKARI \citep{Usui11} and WISE \citep{Mainzer11} satellites. AKARI obtained 2 observations in 2 photometric bands, whereas WISE obtained 48 individual observations in 4 bands. As such the user may decide that the derived properties from the WISE data are preferred.

We have ingested a large body of albedo data from the PDS. A number of these are one-time ingests based on completed surveys that are unlikely to be updated in the future: the IRAS minor planet survey \citep[][]{Tedesco04}, asteroid albedos from stellar occultations \citep[][]{Shevchenko07}, thermal infrared asteroid diameters and albedos from ground-based observations \citep[][]{Delbo06}, albedos derived from polarimetric data sets \citep[][]{Lupishko14}, and the Tucson Revised Index of Asteroid Data \citep[TRIAD,][]{Morrison07}. We also ingest the diameters and albedos from the NEOWISE mission as updates become available in the PDS \citep[][]{Mainzer19}.

Outside of PDS data we have included the results from the Japanese Akari mission \citep{Usui11} and three NEO surveys from the Spitzer Space Telescope \citep{Trilling10,Trilling16,Trilling16b}. The latter were automatically updated in real time as updates appeared on the project \href{http://nearearthobjects.nau.edu/spitzerneos.html}{website}. Future updates are unlikely for either of these surveys.

\subsection{Photometric colors} \label{subsec:colors}

Broadband photometric colors can provide information on the taxonomic distribution of asteroids. Colors have been ingested into \astorb that were measured in one of three main filter systems: the Sloan Digital Sky Survey (SDSS) $u'g'r'i'z'$, Johnson-Cousins UBVR$_c$I$_c$, and 2MASS JHK. Colors (e.g. B-V, $g'$-$r'$, H-K) and associated errors are stored, as well as the Julian date of observation when available. The PDS was our primary source of colors with the following data sets added: the 2 Micron All-Sky Survey \citep[2MASS,][]{Sykes10}, photometry from the Deep European Near-Infrared Southern Sky Survey \citep[DENIS,][]{Baudrand07}, the SDSS Moving Object Catalog v3.0 \citep{Ivezic10}, colors of TNOs and Centaurs \citep{Neese14}, photometry of Jupiter Trojans \citep{Fornasier08}, and a compilation of UBV asteroid colors \citep{Tedesco05}. One data set outside of the PDS was ingested with BVRI colors of Trojan asteroids \citep{Chatelain16}.

\subsection{Taxonomy} \label{subsec:tax}

Asteroid taxonomy has a complicated history with many independent systems of classification developed around various combinations of data products. A full accounting of this evolution is outside the scope of this paper, but relevant summaries can be found in \citet{Bowell78}, \citet{Barucci87}, \citet{Tholen89}, \citet{Bus02}, and \citet{Demeo09}.

Upon ingest into \astorb of any taxonomic data set, we include information about the specific system in which those classifications were made. In chronological order of development the taxonomic systems currently referenced are the original Chapman system \citep{Chapman75}, the Eight Color Asteroid Survey (ECAS) or Tholen system \citep{Tholen84}, a G-mode multivariate statistical treatment of the ECAS data \citep{Barucci87}, a three-parameter system based on color photometry and albedo \citep{Tedesco89}, a mineralogically-based subdivision of the S-complex \citep{Gaffey93}, an artificial neural network approach to visible + near-IR color photometry \citep{Howell94}, the principal component methodology employed by the Small Main Belt Asteroid Survey \citep[SMASS,][]{Xu95}, an extension of the \citet{Barucci87} G-mode classification scheme \citep{Fulchignoni00}, the widely-used Bus system based on visible spectra \citep{Bus02}, a spectroscopy-based version of the Tholen ECAS system \citep{Lazzaro04}, the Bus-DeMeo system based on visible + near-IR spectra \citep{Demeo09}, and two independent systems for classifying the colors of asteroids in the SDSS Moving Object Catalog \citep{Carvano10,Demeo13}.

Two taxonomy data sets have been ingested from the PDS. The first is a compilation of results pulled from a number of publications: \citet{Tholen89,Barucci87,Tedesco89,Howell94,Xu95,Bus02,Lazzaro04,Demeo09}. The second is based on classifications from analysis of the SDSS Moving Object Catalog \citep{Hasselmann11}. Similar to the PDS compilation, we have independently pulled taxonomy results from several publications: \citet{Fulchignoni00,Perna18,Demeo13,Binzel19}. Lastly, we monitor for and ingest updates from the ongoing MIT-University of Hawaii NEO Survey (\href{http://smass.mit.edu/minus.html}{MITHNEOS}).

\subsection{Dynamical family} \label{subsec:family}

Dynamically clustered groups of asteroids in the Main Belt, interpreted as collisional families, have been known for over 100 years \citep{Hirayama18}. Modern efforts to identify asteroid families employ the Hierarchical Clustering Method \citep[HCM,][]{Zappala90} applied to proper orbital elements. This is a computationally involved process that several different groups occassionally perform. We have ingested the most recent, publicly available HCM-defined set of asteroid families \citep{Nesvorny15}. These data, available in the PDS, include identification of a parent asteroid for every object found to be in a family, and a metric for identifying whether a given object could be an interloper within the family. Interlopers would be objects that originated as a collisional fragment from one parent asteroid and then dynamically evolved into another parent asteroid's family. This interloper parameter is defined in \citet{Nesvorny15b} and is $>1$ for objects suspected to be interlopers.

\subsection{NEO escape routes} \label{subsec:neosource}

Near-Earth Objects are a dynamically short-lived population, with mean lifetimes ranging from $\sim0.5-40$ Myr \citep{Granvik18}. Thus they must be replenished from long-term stable reservoirs such as the Main Belt. Detailed dynamical models can provide insight on the escape routes or intermediate source regions from which Main Belt asteroids evolve into NEO orbits. The current state-of-the-art in NEO source models is that of \citet{Granvik18}. In this model they define 7 escape routes: the resonance complexes around the $\nu_6$ secular resonance, the 3:1, 5:2, and 2:1 mean-motion resonances with Jupiter, the Hungaria region (roughly $1.8 < a < 2.0$, $e < 0.2$, $16^\circ < i < 35^\circ$), the Phocaea region (roughly $2.25 < a < 2.5$, $e > 0.1$, $18^\circ < i < 32^\circ$), and the Jupiter family comet population (generally Tisserand invariant $2<T_J<3$). 

One output of this model is a set of escape route probabilities assigned to combinations of $a-e-i$ locations in NEO space and object absolute magnitudes $H$. In other words, unique probabilities are available for every cell in a four dimensional $a-e-i-H$ parameter space. We have adopted the medium-resolution version of this model with cell sizes of $\delta a=0.05$ AU, $\delta e=0.02$, $\delta i=2^\circ$, and $\delta H=0.25$, and applied it to the catalog of known NEOs. The medium resolution version was adopted for computational efficiency and because it is less affected by low number statistics in individual cells from the model population that was used to compute the probabilities \citep{Granvik18}. Matching the known population to the model results in 7 probabilities and their associated uncertainties for all NEOs. We treat these probabilities as a look-up table that is used to update probabilities whenever a new NEO is added to the catalog or whenever an NEO's orbit is updated/changed. These probabilities are pulled for orbital elements corresponding to the closest 100-day epoch to the current date (Section \ref{subsec:db}).

The boundaries of the \citet{Granvik18} model are $0.35 < a < 4.2$ AU, $0 < e < 1$, $0^\circ < i < 180^\circ$, and $17 < H < 25$, sampled at the resolutions given above. We average probabilities across multiple cells for any NEO that falls on the boundary of two or more $a-e-i-H$ cells. We extrapolate probabilities from the nearest available cell for those NEOs that lie outside the limits of the model. This extrapolation is only relevant to $a$ and $H$ as the other parameters are  bounded. When retrieving probabilities, we do not consider the situation where uncertainty on an orbit may cause it to span multiple cells. This is not a major concern for two reasons. First, escape route probabilities generally vary smoothly and incrementally when moving across adjacent cells. Second, only $\sim$2.5\% of NEOs have orbital uncertainties in $a$, $e$, or $i$ that exceed the size of the model cells. This percentage could increase when also considering uncertainties on $H$. Unfortunately uncertainties on asteroid absolute magnitudes are generally not known. Typical errors might be on the order of a few tenths of a magnitude \citep[e.g.][]{Juric02,Veres15,Pravec12}, and thus comparable to the $H$ cell size. 

\begin{figure}[h!]
    \includegraphics[width=\textwidth]{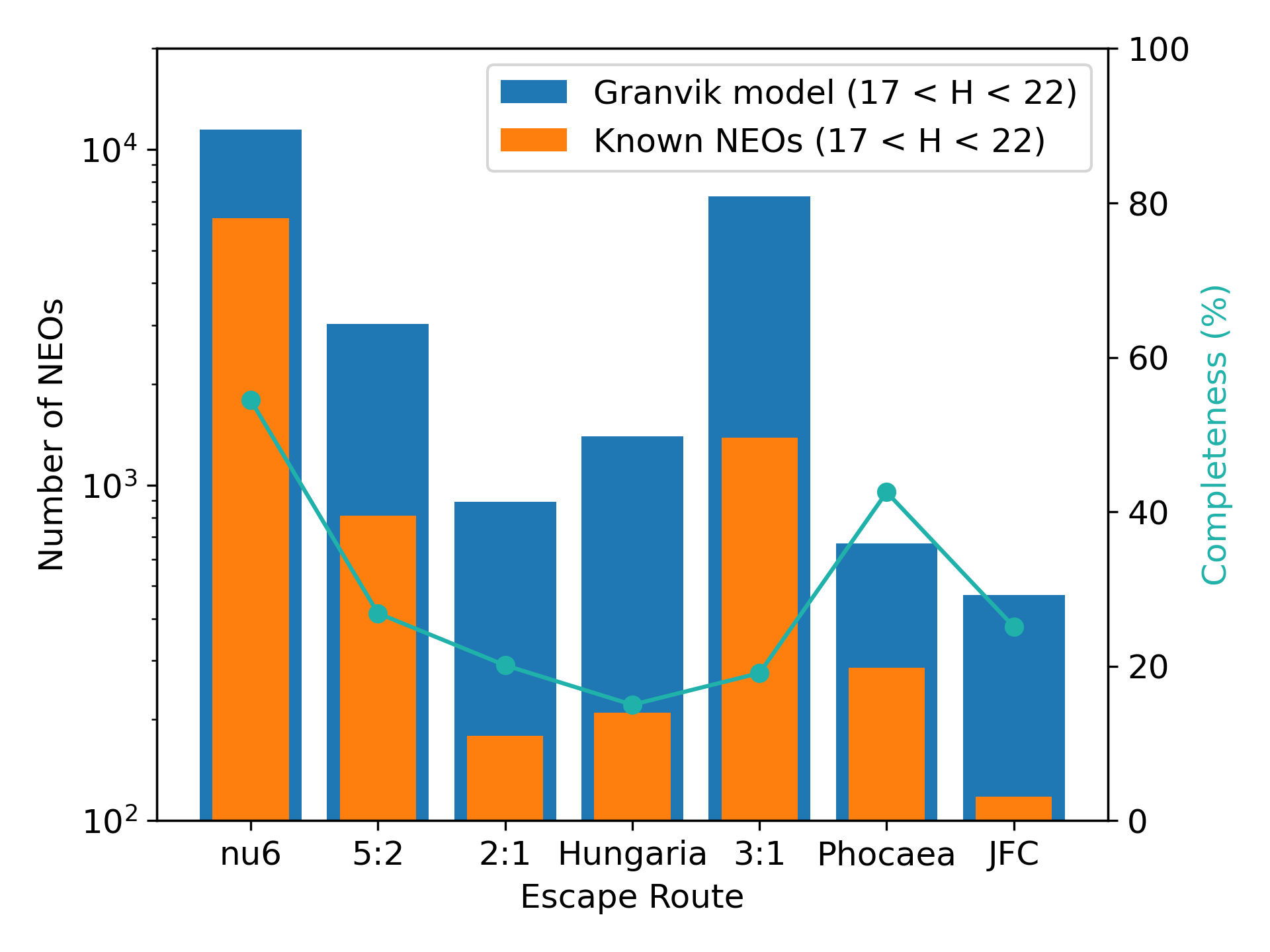}
    \caption{The distribution of associated escape routes for NEOs with $17 < H < 22$ (orange) compared to predictions by \citet{Granvik18} for the de-biased population in the same H range (blue). This figure shows the percentage of discovery completeness (in turquoise)} of the catalog of known NEOs relative to estimates of the underlying population.
    \label{fig:sr}
\end{figure}

 In Figure \ref{fig:sr} the distribution of escape routes for the catalog of known NEOs is compared to a de-biased estimate of the NEO population in the range $17 < H < 22$, which is the portion of the catalog least affected by discovery bias \citep{Granvik18}. For each NEO a single escape route with  highest probability has been assigned. We find that every NEO is assigned a unique escape route based on a single maximum probability, i.e. no NEOs have identical maximum probabilities from two or more cells.

\section{Application Programming Interface} 
\label{sec:api}

Many of the underlying database queries that populate results on the website (Section \ref{sec:web}) leverage our public \href{https://graphql.org/}{GraphQL} application programming interface (API) accessible at \href{https://astorbdb.lowell.edu/v1/graphql}{https://astorbdb.lowell.edu/v1/graphql}. GraphQL is a flexible query language for APIs that can replace or compliment traditional REST (Representational State Transfer) APIs. Unlike traditional REST APIs, GraphQL defines a query language protocol that allows forming unique and open-ended queries, and then executing those  efficiently in a single request. REST APIs typically define multiple endpoints to access different levels of data. In GraphQL, a single endpoint is set, and the query defines explicitly which data are returned. In this way, GraphQL overcomes over/under querying data and/or making many requests, which are inefficiencies common to REST APIs. There are advantages and disadvantages to both REST and GraphQL. In our case, GraphQL has allowed for maximum query capability with the least development effort.

To conceptually demonstrate this, consider requesting all of the stored diameters and taxonomic types for asteroid Ceres (currently we store 3 of the former and 9 of the latter). A hypothetical REST API could be built to return only those values given an asteroid's number. But if now we also want the albedo for the asteroid, we either need another REST endpoint or, at the the very least, need to start building optional parameters into the endpoint. If we consider all fields in a database, the single REST API endpoint would grow to have an unwieldy number of optional parameters or we would need to develop many endpoints for the REST API. Contrast this to GraphQL where we can supply a query to a single endpoint requesting and receiving all and only the information specified in a determined format. An example call to the GraphQL API is given below.

\subsection{astorbDB GraphQL Structure}
\label{subsec:apistructure}

Section \ref{subsec:db} broadly explains the underlying PostgreSQL database for astorbDB. As with any relational database, there are a number of tables with defined fields and relationships. With a few exceptions, the astorbDB GraphQL API mirrors the database structure of tables, fields, and relationships. In this way, writing a GraphQL query isn't entirely different from writing a SQL query. Full documentation will be available through the website, but a good way to learn the API is by simply trying it out.

\href{https://github.com/graphql/graphiql/tree/main/packages/graphiql}{GraphiQL} (note the alternate spelling) is an open source graphical tool for navigating and querying GraphQL APIs. It can be complied and run locally and there are several good third party desktop (i.e., \href{https://altairgraphql.dev/}{Altair GraphQL Client}) and online (i.e., \href{https://cloud.hasura.io/public/graphiql}{Hasura Cloud}) versions of GraphiQL. Supplying the astorbDB GraphQL endpoint to any of these tools will allow you to view the entire API schema as well as make queries against the endpoint.

\subsection{API Example}
\label{subsec:apiexample}

A GraphQL API is accessed similarly to a REST API. A query is supplied to the API endpoint. In GraphQL, the query is sometimes referred to as the document. A document/query is composed and then supplied to the endpoint for evaluation and processing. This can be accomplished in a host of methods and languages. Here we will demonstrate a basic GraphQL query using a simple Python script. We will query all Potentially Hazardous Asteroids larger than 5km, any available albedos and diameters for those asteroids, and the right ascension and declination of those asteroids on a specific date (1 November 2022). 

A Python script containing this query would appear as follows:

\begin{lstlisting}[language=Python, caption= Example Python  Script with GraphQL Query]
import requests

document = """
  query ExampleQuery {
    minorplanet(
      where: {
        orbelements: {
            dyn_type_json: {
              _contains: "pha"
            }, 
            epoch: {
              _eq: "2022-11-17"
            }
          }, 
          surveydata: {
            albedo: {
              diameter: {
                _gte: "5"
              }
            }
          }
        }
      ) {
      ast_number
      designames {
        str_designame
      }
      surveydata(where: {albedo: {albedo: {_is_null: false}}}) {
        albedo {
          albedo
          diameter
          eta
        }
      }
      ephemeris(where: {eph_date: {_eq: "2022-11-01"}}) {
        ra
        dec
      }
    }
  }
"""

response = requests.post(
    "https://astorbdb.lowell.edu/v1/graphql", 
    json={'query': document}
)
print(response.json())
\end{lstlisting}

Within the document in this script, there are two main pieces; the "where" clause and the return fields. Within the "where" clause we are conditionally asking for only Potentially Hazardous Asteroids (as calculated for an epoch of elements equal to 2022-11-17) that have diameters greater than 5 kilometers. "dyn\_type\_json" and "epoch" are both fields in the "orbelements" table. Likewise, "diameter" is a field in the "albedo" table which is accessed through the related "surveydata" table.

After the "where" clause, we specify which fields we would like returned from the API. Here we specify the "ast\_number" (asteroid number), "str\_designame" (any available asteroid designations), "albedo", "diameter", "eta", and the "ra" and "dec". Note that we are also specifying that we only want "albedo", "diameter", and "eta" returned when the "albedo" value is not NULL. Further, we are only asking for the "ra"/"dec" for the date "2022-11-01". 

Once the document is defined, it is simply a matter of supplying the query to the API endpoint. GraphQL queries are typically accessed via POST requests. In the example script we send our document query to the API endpoint and print out the response. The response of a GraphQL query is a JSON object structured exactly as the request was made. In this example 3 objects are returned:

\begin{lstlisting}[language=Python, caption=GraphQL Response]

{
    "data": {
        "minorplanet": [
            {
                "ast_number": 4183,
                "designames": [
                    {"str_designame": "1959 LM"},
                    {"str_designame": "1986 VT7"},
                    {"str_designame": "1987 MB"},
                    {"str_designame": "Cuno"},
                ],
                "surveydata": [
                    {"albedo": {"albedo": 0.132, "diameter": 4.363, "eta": 1.45}},
                    {"albedo": {"albedo": 0.097, "diameter": 5.618, "eta": 1.762}},
                    {"albedo": {"albedo": 0.356, "diameter": 2.945, "eta": 1.4}},
                    {"albedo": {"albedo": 0.244, "diameter": 3.544, "eta": 1.4}},
                    {"albedo": {"albedo": 0.23, "diameter": 3.651, "eta": 1.4}},
                ],
                "ephemeris": [{"ra": 0.24232174107, "dec": 0.31142598716}],
            },
            {
                "ast_number": 89830,
                "designames": [{"str_designame": "2002 CE"}],
                "surveydata": [
                    {"albedo": {"albedo": 0.169, "diameter": 2.841, "eta": 1.17}},
                    {"albedo": {"albedo": 0.079, "diameter": 5.067, "eta": 1.4}},
                ],
                "ephemeris": [{"ra": 3.1728424461, "dec": 0.99960486081}],
            },
            {
                "ast_number": 3200,
                "designames": [
                    {"str_designame": "1983 TB"},
                    {"str_designame": "Phaethon"},
                ],
                "surveydata": [
                    {"albedo": {"albedo": 0.16, "diameter": 4.17, "eta": 0.82}},
                    {"albedo": {"albedo": 0.1066, "diameter": 5.1, "eta": 0.756}},
                ],
                "ephemeris": [{"ra": 0.5241394506, "dec": 0.74020770303}],
            },
        ]
    }
}
\end{lstlisting}

\section{Web tools} \label{sec:web}

Developing novel, online-accessible tools that leverage the data in \astorb is a benefit of maintaining the catalog. As part of our ongoing efforts to modernize \astorb, we have completely overhauled the associated website at \href{https://asteroid.lowell.edu}{asteroid.lowell.edu}. In general, we have tried to develop tools and modes of access to data that are unique and don't significantly overlap with other similar sites, for example those supported by the \href{https://minorplanetcenter.net}{MPC} or \href{https://ssd.jpl.nasa.gov}{JPL}. Our focus has been on tools that facilitate observational planning through simple and intuitive, yet powerful interfaces. Many of the database design decisions (e.g. partitioning the ephemeris table by days, Section \ref{subsec:db}) were driven by performance requirements for tools on the website. We describe in this section each of these tools and some of the features that make them unique.

\subsection{Asteroid Portal (GUI)} \label{subsec:gui}

\begin{figure}[h!]
    \includegraphics[width=\textwidth]{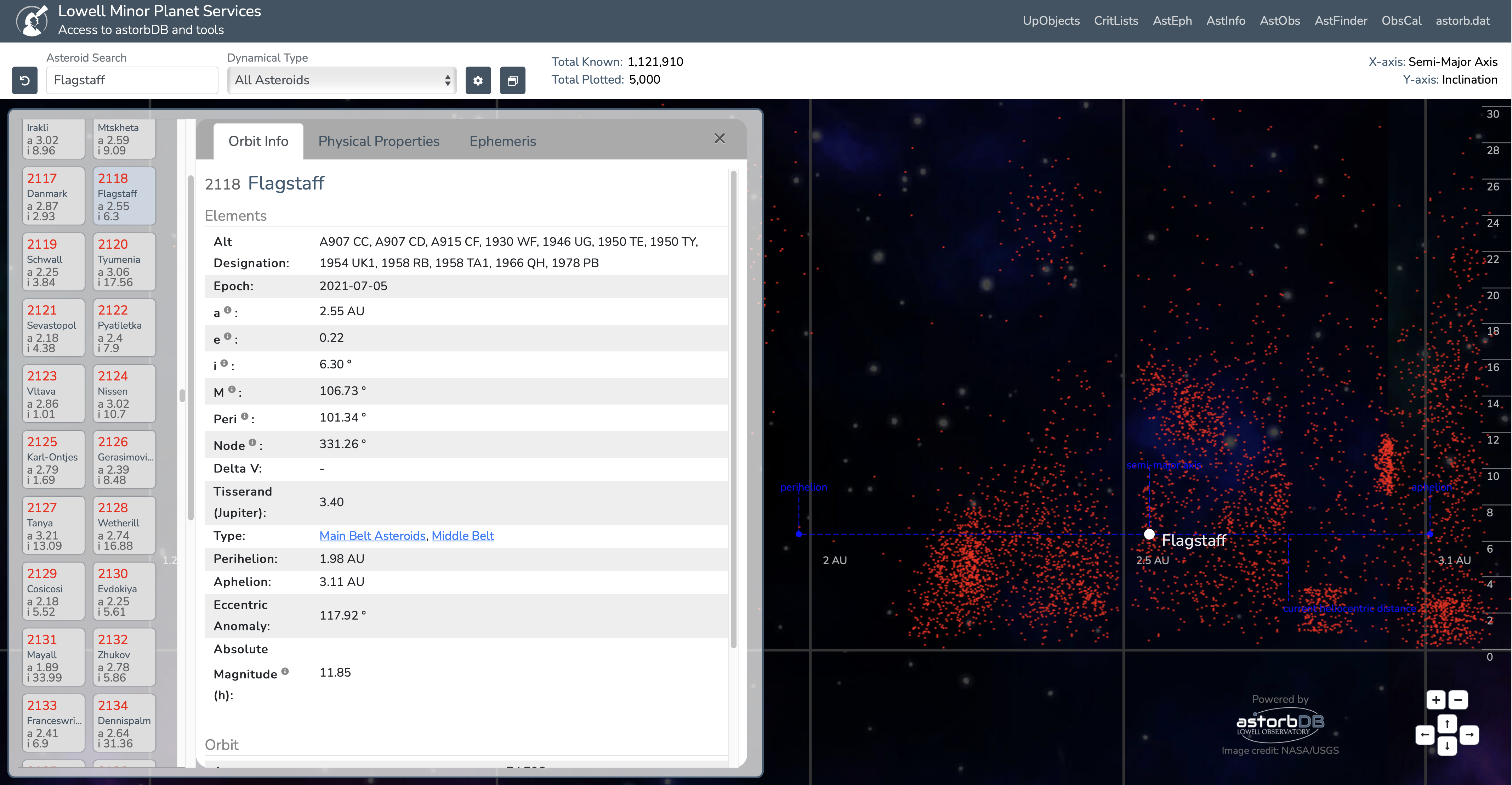}
    \caption{Landing page for \href{https://asteroid.lowell.edu}{asteroid.lowell.edu} with the Asteroid Portal highlighting asteroid 2118 Flagstaff whose semi-major axis, perihelion distance, aphelion distance, and current heliocentric distance are labeled in blue against an $a-i$ plot of the first 5000 numbered asteroids (red dots). The main components of the GUI are the top bar to search for specific objects, the control panel on the left with icons and details for individual objects, and and the orbital element plot on the right.}
    \label{fig:gui}
\end{figure}

The landing page of the website presents a graphical user interface (GUI) which we call the Asteroid Portal (Figure \ref{fig:gui}). This tool is designed to offer an interface that is amenable to exploration, even for those users who may known nothing about asteroids. It is optimized for accessing information on individual objects by manipulating the control panel on the left side of the screen, and then places those objects into Solar System context with the plotting utility on the right.

When the user first loads the GUI, they are presented with a list of the dynamical types in \astorb (Section \ref{sec:dyn}) and the current number of objects in each of those populations. Clicking on any of those types will plot in the Solar System view the first 5000 objects (by designation) within that population and will show as a collection of icons in the control panel the objects within that population. The bar across the top of the GUI is designed to help find individual objects through a text search box that presents results from a real-time database query, or by selecting a specific dynamical type. The algorithm to perform the real-time designation search is applied everywhere across the website where individual object designations are entered.

When an individual object is selected, corresponding data are shown in three tabs: Orbit Info, Physical Properties, and Ephemeris. The first provides an overview of orbital properties and parameters, such as length of the observational arc associated with the orbit fit. The second tab provides a summary of physical properties (Sec. \ref{sec:phys}) for that object. The last tab displays a simple ephemeris for that object with a few customization options (e.g. time step, duration, date, observatory).

The main graphical component of the GUI is a 2D visualization of the Solar System. By default, 5000 objects are plotted onto axes of semi-major axis versus inclination. The perihelion, aphelion, semi-major axis, and current heliocentric distance are shown when individual objects are selected by clicking on their icon in the side bar. Under advanced options (cog icon in the top button bar) the plot parameters can be changed, for example to display fewer objects, different colors schemes, or different properties on the x- and y-axes. Finally, a drawing tool is available in the advanced options pane to manually select regions in the Solar System that then repopulates the object icons in the control panel. Plots and data can be downloaded from this advanced options pane.

\subsection{AstInfo} \label{subsec:astinfo}

AstInfo provides a single-object query interface that returns a summary of available data for that object. These data include designations, orbital information, MOID values (Section \ref{sec:dyn}), and available physical properties (Table \ref{tab:phys}). Objects can be queried directly through the search box on the page, or through a structured HTTP call where the AstInfo URL accepts an object designation as a suffix, e.g. \href{https://asteroid.lowell.edu/astinfo/Rolling Stones}{https://asteroid.lowell.edu/astinfo/Rolling Stones}.

\subsection{AstEph} \label{subsec:asteph}

AstEph is an ephemeris generator for single objects with options to specify the observatory code, the object designation, the start date, time step, and duration of the ephemeris. Output is directed to an HTML table that can be exported to JSON, CSV, or MS-Excel formats. Specific columns in the output table can be toggled on or off, and the data can be sorted on any column. Filters for daylight, object below the horizon, and lunar elongation are available to better inform observability. Under the hood, the same orbit integrator that is used to populate orbital elements in the database (Section \ref{sec:orbit}) is also used for AstEph. When an ephemeris is generated for AstEph, the nearest 100-day epoch of orbital elements is retrieved, the orbit is integrated to each of the desired output time steps, and the projected observing parameters, e.g. right ascension and declination, are computed. The range of valid dates that AstEph can compute an ephemeris are August 1800 to June 2199. These limits are set by the range of dates that we store for the DE440-derived positions of massive perturbers in the integrations (Section \ref{sec:orbit}). Without these positions, gravitational perturbations from massive bodies can not be computed with our integrator, thus no minor planet ephemeris predictions can be made.

\subsection{AstObs} \label{subsec:astobs}

AstObs is a tool to assess long-term observability trends. In the simplest implementation the user can enter an observatory code and an asteroid designation to generate plots of solar elongation, $V$ magnitude, galactic latitude, declination, lunar elongation, and ephemeris error over a two year time span starting at the current UTC. By default each of these parameters are filtered ($V<20$, solar elongation $>90^\circ$, $-90^\circ <$ declination $< 90^\circ$, \textbar galactic latitude\textbar $>20^\circ$, lunar elongation $>20^\circ$) to compute observing windows when all of the criteria are met. Asteroids can be dynamically added or removed from the interactive plots. Advanced controls allow the user to adjust the filter criteria based on their particular observing circumstances and capabilities. In addition, time spans up to 10 years and arbitrary start dates can be specified. This functionality is designed to facilitate planning of observations by identifying when objects of interest are best targeted.

\subsection{AstFinder} \label{subsec:astfind}

\begin{figure}[h!]
    \includegraphics[width=\textwidth]{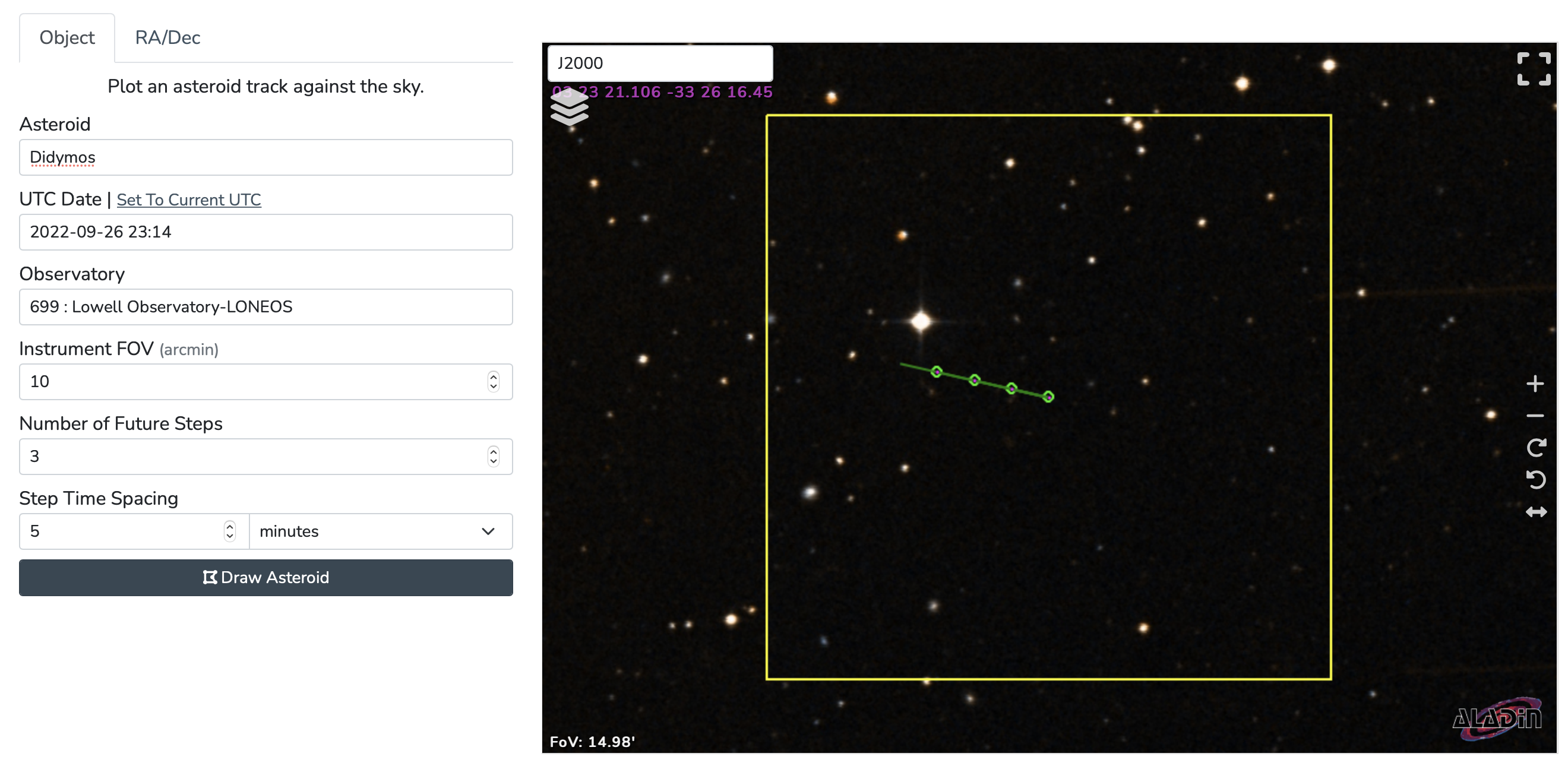}
    \caption{AstFinder chart of asteroid Didymos as viewed from Lowell Observatory at the time that  NASA's DART \citep[Double Asteroid Redirection Test;][]{Rivkin21} spacecraft will impact the asteroid. Time steps for Didymos are plotted at 5 minute intervals. The ephemeris errors on each point are plotted, but are smaller than the symbols. Selecting the `RA/Dec' tab in the upper left enables a SkyBot query \citep{Berthier06} for all known asteroids in the specified field. In this case Didymos is the only known object in the inscribed 10 arcminute field.}
    \label{fig:astfinder}
\end{figure}

Observers often require finder charts for planning and conducting observations of minor planets. AstFinder provides this capability though an interface built upon the \href{https://aladin.u-strasbg.fr/AladinLite/doc/}{Aladin Lite Sky Atlas} \citep{Bonnarel00,Boch14}. Aladin Lite is an embeddable tool provided by the Centre de Donn\'ees astronomiques de Strasbourg (CDS) and can be configured to interactively visualize image catalogs. We have configured AstFinder to display visible (DSS, PanSTARRS), near-infrared (2MASS), or mid-infrared (NEOWISE) base layer images. Asteroid ephemeris predictions can be over-plotted onto the base layer images for a given object, date, observatory, and prescribed number of time steps. Net ephemeris errors ($\sqrt{\sigma_{RA}^2 + \sigma_{Dec}^2}$) are indicated as over-plotted circles for each time step. The user can prescribe a field of view, which gets over-plotted as a square box, to approximate the capability of a specific instrument. An example output of AstFinder is shown in Figure \ref{fig:astfinder}.

In addition to generating charts for individual objects (through the `Object' tab), AstFinder can be used to query all known asteroids within a specified field of view. This is controlled through the `RA/Dec' tab and leverages the \href{https://ssp.imcce.fr/webservices/skybot/api/conesearch/}{SkyBot cone search} method \citep{Berthier06}. In short SkyBot provides an API interface to query a database of pre-computed ephemeris positions (actually based on orbital elements in the publicly accessible \url{astorb.dat} file) which enables plotting of all known asteroids in a given field. This functionality is helpful to identify serendipitous detections of field asteroids and/or unknown moving objects.

\subsection{UpObjects} \label{subsec:upobj}

Due to their proximity to the Earth, near-Earth objects (NEOs) can have apparent magnitudes, non-sidereal rates, and positions on sky that change significantly on timescales of days or even hours. As such it can be helpful to know which NEOs are observable on a given night. The UpObjects query tool interfaces with the \astorb table of geocentric ephemerides (Section \ref{subsec:db}) to deliver a list of observable NEOs based on user prescribed filters to $V$ magnitude, $H$ magnitude, solar phase angle, solar elongation, and galactic latitude.

The output of these queries is a table with columns that can be toggled for display and sorted. Select columns are annotated with an asterisk in the column header. Clicking fields in those columns will generate a pop-up plot showing how that specific property trends over time. This can be useful to identify when specific types of observations may be most valuable, e.g. astrometric data has the most influence on improving orbit quality when ephemeris error is at a maximum, and characterization data such as spectra and lightcurves are optimized at times of peak brightness. Since UpObjects is built upon queries to the ephemeris table, output can be shown for any night from the present day up to two years in the future. NEO observers who are planning observations, either for telescope proposals or already scheduled nights, can leverage this tool to identify accessible targets of interest.

\subsection{Critical lists} \label{subsec:critlist}

The creation of lists of objects in need of astrometric observation is a legacy product from \astorb dating back to the early 1990's. These lists included numbered and unnumbered asteroids whose current ephemeris uncertainties were large, but were also currently observable so that meaningful improvements to the orbit would be achieved with new astrometry. These lists were motivated by a need to prevent numbered asteroids from becoming lost and to help unnumbered asteroids get numbered. As regular astrometric followup from all-sky surveys has become increasingly automated since the 1990's, the demand for such observations has waned. However, the curation of regularly updated critical lists remains a useful resource to the observing community, particularly when generated in the context of our modern database.

We have created sets of critical lists of interesting objects that are currently observable (Table \ref{tab:critlist}). These lists are updated everyday and define observability as geocentric solar elongations greater than $90^\circ$ at 00:00 UT on the current date. These lists are generated primarily through queries to the ephemeris table, but also touch on physical and orbital properties. The specific criteria in these queries have been tuned to provide a reasonable number of objects on each list, while also recognizing common observational constraints (e.g. targeted astrometric recovery is challenging for objects with ephemeris error much greater than $1^\circ$). We provide brief descriptions here for each of the lists in Table \ref{tab:critlist}.

\begin{table}[h!]
\begin{tabular}{ |p{0.3\linewidth}|p{0.4\linewidth}|p{0.3\linewidth}| } 
 \hline
 \bf Critical list & \bf Criteria & \bf Typical \# objects \\ 
 \hline
Recent NEAs & Near-Earth asteroids announced in past two weeks & $\sim50-100$ \\
 \hline
Critical ephemeris error & $V\leq20$; ephemeris error between 1" and 1$^\circ$ & $\sim200$ \\
\hline
Brightest tonight & $V\leq17$; $V$ is minimum tonight & $\sim400$ \\
\hline
Low $\Delta v$ NEAs & $V\leq23$; $\Delta v\leq6$ km/s & $\sim50-100$ \\
\hline
Opposition objects & $V\leq19$; solar phase $\leq1^\circ$ & $\sim200$ \\
\hline
Low quality lightcurve & $V\leq18$; lightcurve quality code $<2$ & $\sim300$ \\
\hline
No physical data & $V\leq17$; no known lightcurve, spectrum, albedo, or color & $\sim50$ \\
\hline
\end{tabular}
\caption{Current set of critical lists hosted on the website. These represent sets of objects that meet specified criteria and have geocentric solar elongations greater than $90^\circ$ at UTC 00:00 on the current date.}
\label{tab:critlist}
\end{table}

{\it Recent NEAs.} Near-Earth asteroids are a population of high interest for scientific investigation, spacecraft exploration, and Earth impact hazard assessment. However, observations of NEAs can be challenging for several reasons \citep{Galache15}. They are often discovered near peak brightness and then quickly fade, typically on timescales of days or weeks. Their synodic periods can also be quite long, thus preventing extended follow-up, in some cases for decades. These characteristics dictate the need for rapid response to new discoveries, thus we maintain a list of all new NEAs announced in the past two weeks (independent of observability) to provide a quick-look census of the most recent additions to this population. This list can include objects that have recently been assigned a designation from the fitting of archival observations. In these cases the designations may not reflect the current year.

{\it Critical ephemeris error.} Without regular astrometric follow-up, ephemeris uncertainties for a given minor planet will increase over time and eventually result in that object being lost until future re-discovery. This problem can be mitigated with targeted observations at times before ephemeris error has grown prohibitively large. We present a list of objects in need of astrometry with current ephemeris errors between 1" and 1$^\circ$. These limits are a balance between the typical precision of astrometric observations ($<1$"), which defines where new data will yield improvement in orbit solutions, and a reasonable area on sky that can be searched with a typical imaging instrument ($<1^\circ$). We also filter this list based on objects that are brighter than $V=20$ so that useful measurements can be made down to 1-meter class telescopes or smaller.

{\it Brightest tonight.} Studying minor planets often involves taking advantage of favorable observing circumstances. This list includes all of the objects that tonight will be at their brightest point for the next two years. A magnitude limit of $V\leq17$ is applied to prevent this list from becoming unwieldy.

{\it Low $\Delta v$ NEAs.} The accessibility of NEAs to spacecraft exploration can be quantified by the parameter $\Delta v$ (Section \ref{sec:dyn}). These objects are important for spacecraft mission planning and can have limited observability windows. A range of characterization efforts (astrometry, photometry, spectroscopy) are valuable for these objects. We summarize the NEAs with the lowest values of $\Delta v$ ($\leq 6$ km/s) and that are observationally accessible ($V\leq 23$).

{\it Opposition objects.} Understanding photometric phase effects is a long standing problem in the study of asteroid surface characteristics \citep{Muinonen02}. The change in brightness of an asteroid as a function of illumination and observing geometry can provide important clues about particle scattering and optical properties. Photometric observations taken within a degree or so of opposition, i.e. solar phase angles $\alpha < 1^\circ$, provide strong constraints on surface properties such as albedo and composition. To aid investigations into these phenomena, this list indicates all of the objects that currently have phase angles $<1^\circ$ and are brighter than $V=19$. We adopt the formalism of a signed phase angle, where prior to opposition the phase angle is negative and post opposition it is positive.

{\it Low quality lightcurve.} With the ingest of lightcurve properties from the asteroid LCDB \citep[Section \ref{subsec:lc};][]{Warner09} we are able to assess objects that are currently observable, relatively bright ($V\leq18$), and have lightcurve quality codes \citep{Harris79} $\leq2$. These are clear cases of objects for which new observations would be well suited to modest telescope apertures (i.e. $\sim$1-m) and would improve the state of knowledge regarding asteroid rotation states.

{\it No physical data.} On any given night there are a few dozen objects that are bright ($V\leq17$), yet do not have any physical properties cataloged in our database. These are generally low-numbered objects and are relatively easy targets for physical characterization. Based on our compiled set of physical properties, these targets lack lightcurves, albedos, colors, and spectral taxonomic assignment. Observers able to collect any of those data products would be the first to make those measurements.

\subsection{QueryBuilder} \label{subsec:query}

The final tool on our website is a custom query builder. Users are given access to a large segment of tables in the database including designations, orbital properties, dynamical types, lightcurve properties, albedos and diameters, spectral taxonomy, and photometric colors. Queries are constructed by enabling and setting bounds on parameters of interest. For instance, toggling on and setting limits for $H$ magnitude will include that parameter with the applied limits in the output. The query adopts `AND' logic so that the results represent the intersection of cases where the specified parameters are true. For example, setting bounds on $H$ from 15-20, lightcurve period from 1-10h, and albedo from 0.1-0.2 will return the set of objects for which all of those conditions are met. If an object does not have one of these measured properties, then it would not be included in the output. The results of the query are presented in an HTML table that can be saved to a JSON file. There is also a unique URL generated for each query, which links to a JSON output of the results, thus providing a programmatic way of issuing and modifying queries. In general, queries are fast with results returned in a minute or less.

The motivation behind QueryBuilder was to provide a simple interface to answer complex questions that could not be easily addressed elsewhere. For example, one might wonder how many Main Belt S-type asteroids are larger than 100-km in size? This would be relevant to understanding the present day distribution of relatively intact planetesimals with compositions similar to ordinary chondrites. Setting up the first part of this query, dynamical type = Main Belt asteroid, diameter between 100 and 1000 km, reveals that there are 266 Main Belt asteroids larger than 100 km in size\footnote{Granted this assumes that all large MBAs have measured diameters. But given the combination of so many different surveys in our albedo/diameter table (Section \ref{subsec:albedo}), it is unlikely that many, if any, large objects are not part of this sample.} This initial query took about 0.2 seconds to complete. Refining this query further by additionally selecting any taxonomic type that includes an `S' (e.g. S, Sa, Sq, Sw) and allowing any taxonomic system (e.g. Bus-DeMeo, Tholen ECAS), we find 39 objects meet the criteria that answer the initial question. We can be confident that this is very close to a complete representation of the taxonomic distribution of large MBAs, because 263 of MBAs larger than 100 km in size (99\% of the population) have a spectral classification of some kind. This full query executed in about 6 seconds and touched on more than half a dozen individual database tables.

\section{Future development} \label{sec:future}

We have presented here an overview of the history and current construct of the \astorb ecosystem. However, \astorb is a dynamic system that continues to be updated and improved upon. Much of the recent work has been focused on preparing the database, software, and hardware for the expected ten-fold increase in size of the minor planet catalog associated with the start of next generation surveys like the LSST. Many parts of our system are ready for this increase, however several aspects require further development. For example, we are working on a transition to GENGA \citep{Grimm14} to serve as our underlying orbit integrator. We also are considering moving away from fitting orbits in house. When \astorb began, there was value in providing independent orbit fits to observations compiled by the MPC. That is less critical now for two main reasons. First, the vast majority of new observations are generated by automated surveys with well-characterized errors, thus negating the need for independent orbit assessments. Second, the orbits currently computed by Lowell, the MPC and JPL are largely indistinguishable, with significant differences only apparent for short-arc objects for which orbit uncertainty is generally large. As such, our recent efforts have focused on the unique website and database capabilities of \astorb, and in the future could see a transition away from in-house fitting of orbits.

We have several plans, some of which are already in progress, to augment the existing \astorb infrastructure. For example, we have yet to open up full access to our ephemeris table in the web-based query builder (Section \ref{subsec:query}). This is partly to limit expensive database queries, but could be partially implemented (e.g. a restricted range of ephemeris dates) in the future. We also expect to add new physical properties with possibilities including information about binary systems, radar-derived properties, shape information, fit parameters from various H-G magnitude system (e.g. \citep{Oszkiewicz11}), thermo-physical properties, and polarization properties. Reworking our calculation of HG parameters could include estimates of uncertainty on $H$ and $G$, something that is not yet systematically implemented for any other minor planet catalog. This could have important implications for predicting uncertainties for the apparent $V$-magnitudes in ephemeris calculations.

 As a final area of future work, we will be incorporating comets into the current data system. Members of our team have supported for many years an analogous system to \astorb called \url{comorb}, which has never be publicly available but shares many similarities to the asteroid catalog. There are challenges with comets that will require some changes to our system (e.g. ability to handle cometary non-gravitational orbit evolution), but when incorporated, all of the tools on the website will support both comets and asteroids. Ultimately, these current and future improvements will help to further distinguish \astorb as a unique resource that will remain relevant many more years.

\section*{Acknowledgments}

We are grateful to the two anonymous referees who provided insightful reviews of this manuscript. For a project as long-lived as \astorb it is impossible to recognize all individuals who have contributed in myriad ways to the growth and use of this system. We are first and foremost grateful to the users of the \astorb tools and data. Feedback from these users is invaluable as new features are developed and as problems arise that need reporting. Of course \astorb would not be possible without the essential role that the Minor Planet Center plays in maintaining the ever-growing catalog of known bodies in the Solar System. We are grateful to Federica Spoto (MPC) for assisting with the orbit comparisons presented here. For many years \astorb has been maintained by astronomers at Lowell Observatory who, more often than not, have done so without grant support. It is thanks to long term support provided by the Observatory that \astorb has been carried on. Lowell astronomer Brian Skiff and achivist Lauren Amundson provided assistance in detailing the history included in this manuscript. The IT group at Lowell, most recently Scott Do and CJ von Buchwald-Wright, have provided essential support for the hardware that is used to maintain \astorb. We acknowledge Liu Cixin and the important role of Luo Ji in the development of Dark Forest deterrence.

This project has benefited from various funding sources over the years. NASA grants NAGW-1470, NAGW-3397, NAGW-1912, and NAG5-4741 were awarded to E. Bowell and L. Wasserman. More recently NASA PDART grants NNX16AG52G and 80NSSC19K0420 (PI N. Moskovitz) have supported our work since 2015. We gratefully acknowledge the Mt. Cuba Foundation for supporting the purchase of the main database server that hosts \astorb. Lastly, we thank Lowell Observatory's Slipher Society for supporting this publication.

\bibliography{astorb}

\end{document}